\documentclass[twocolumn,twocolappendix]{aastex631}
\hypersetup{linkcolor=magenta,citecolor=blue,filecolor=cyan,urlcolor=blue}
\usepackage{graphicx}
\usepackage{amssymb}
\usepackage{amsmath}

\shorttitle{ Constraining Solid HNCO and Its Environment in Open JWST Data with Laboratory Spectra}
\shortauthors{Nakibov R. et al.}
\begin{document}
    \title{
    Constraining Solid HNCO and Its Environment in Open JWST Data with Laboratory Spectra}

\correspondingauthor{Ruslan Nakibov}
\email{nakibov.ruslan@urfu.ru}

\author[0009-0004-5420-8824]{Ruslan Nakibov}
\affiliation{Research Laboratory for Astrochemistry, Ural Federal University, \\ Kuibysheva St. 48 Yekaterinburg 620026, Russia}
\author[0009-0007-0109-3439]{Varvara Karteyeva}
\affiliation{Research Laboratory for Astrochemistry, Ural Federal University, \\ Kuibysheva St. 48 Yekaterinburg 620026, Russia}
\author[0000-0001-5618-4660]{Katerina Borshcheva}
\affiliation{Research Laboratory for Astrochemistry, Ural Federal University, \\ Kuibysheva St. 48 Yekaterinburg 620026, Russia}
\affiliation{Institute of Astronomy of the Russian Academy of Sciences \\ Pyatnitskaya St. 48 Moscow 119017, Russia}
\author[0000-0002-1584-3620]{Igor Petrashkevich}
\affiliation{Research Laboratory for Astrochemistry, Ural Federal University, \\ Kuibysheva St. 48 Yekaterinburg 620026, Russia}
\author[0009-0001-3921-604X]{Mikhail Medvedev}
\affiliation{Research Laboratory for Astrochemistry, Ural Federal University, \\ Kuibysheva St. 48 Yekaterinburg 620026, Russia}
\author[0000-0003-1684-3355]{Anton Vasyunin}
\affiliation{Research Laboratory for Astrochemistry, Ural Federal University, \\ Kuibysheva St. 48 Yekaterinburg 620026, Russia}
\begin{abstract}
{The presence of HNCO in the gaseous phase is well known for the interstellar medium, but the molecule is not detected yet in the solid phase. We report the first laboratory-confirmed identification of solid HNCO at 2260 cm$^{-1}$ in open JWST data using laboratory infrared spectra. For that, we grew binary mixtures of HNCO with main constituents of interstellar ice using ISEAge setup at 10 K: HNCO:H$_2$O = 1:10, HNCO:CO = 1:10, HNCO:CO$_2$ = 1:10, HNCO:CH$_3$OH~=~1:10. We present transmission infrared spectra of pure HNCO and HNCO-bearing ices along with the corresponding band strengths at 10 K. The fitting of open JWST observational data with the laboratory reference spectra in a 10--23 K temperature range show that solid HNCO is predominantly located in the apolar layer rich in CO. The abundance of solid HNCO is approximately estimated as 0.04--0.15\% relative to solid H$_2$O. The positive correlation between HNCO and N$_2$O suggests their co-formation via the chemical route that includes solid N$_2$. The results are also supported by the numerical modeling of HNCO formation in protostellar objects with predicted abundance of 0.06--0.11\%. We calculated the upper limits for CH$_3$CN toward two protostars and evaluated its abundance to solid water using the spectroscopic fitting and modeling.}
\end{abstract}


\section{Introduction} \label{sec:intro}
Isocyanic acid (HNCO) is the simplest molecule that contains both hydrogen and the three key elements of the CNO cycle—carbon, nitrogen, and oxygen—making it relevant for astrobiology and prebiotic chemistry. HNCO is the most stable and the most abundant isomer of the CHNO species \citep{2010ApJ...725.2101Q, 2018MNRAS.474.2796Q}. Also, it is one of the simplest carriers of the so-called `peptide bond' \citep{2011ApJ...743...60H}. Solid HNCO can act as a precursor to `peptide-like' molecules with amide functional group \citep{2018MNRAS.480.3628L}. The presence of gaseous HNCO is confirmed at various stages of star formation, such as cold dense cores and protostars \citep{2009A&A...494L..13O,2017MNRAS.467.2723P,2019MNRAS.483.2014H}. Gaseous HNCO was also detected in interstellar medium \citep{1972ApJ...177..619S, 2022A&A...665A.131L} and comets \citep{2024A&A...690A.271B}. Grain surface origin of HNCO is implied from its enhanced gas-phase abundance in comparison to quiescent clouds in hot cores and corinos \citep{2010A&A...516A.105M,2010A&A...510A..85T} and shocks \citep[e.g.,][]{2017MNRAS.467.2723P,2023ApJ...949...89X}. \cite{2004A&A...415..425V} calculated solid HNCO upper limits relative to water ice for NGC 7538 IRS9 ($\le$0.5$\%$), GL 2136 ($\le$0.7$\%$) and W33A ($\le$0.5$\%$). Finally, \cite{2026AA...706L...4K} presented tentative detection of solid HNCO along with nitrous oxide (N$_2$O) in 13 objects via its 4.42~$\mu$m absorption band \citep[$\nu_2;$][]{2002AcSpA..58...73L,2004A&A...415..425V} using the James Webb Space Telescope Near-Infrared Spectrograph (JWST NIRSpec) open data. Column density values in range of (0.2--2.9)\texttimes10$^{16}$ cm$^{-2}$ were derived from Gaussian fitting of the HNCO feature. However, the shape of ice absorption bands often deviates from a Gaussian profile, as it depends on the molecular environment and ice temperature. Therefore, Gaussian fits should be validated against laboratory reference spectra.

HNCO was included in gas-grain modeling in e.g., \cite{2008ApJ...682..283G,2009A&A...499..215B,2010ApJ...725.2101Q,2017A&A...601A..49B,Garrod_ea2022}. \cite{Garrod_ea2022} reported the solid HNCO estimations for `basic' and `final' models with `medium' warm-up timescale in their Fig. 2 and Fig. 11, respectively (see their Table 1 for the description of `basic' and `final' models). The abundances relative to H$_2$O for solid HNCO at 20 K are 0.4\% for `basic' and 0.05\% for `final' (from their Fig. 2 and Fig. 11, respectively). The difference is likely caused by the uncertainty in CO+OH reaction products branching ratio (see their Section 2.9.4).

Identification of HNCO would clarify the OCN$^{-}$ formation pathways as HNCO is one of the main OCN$^{-}$ precursors \citep{2011A&A...530A..96T}. \cite{2004A&A...415..425V} assessed the OCN$^{-}$ formation in mixtures with H$_2$O and NH$_3$ under UV-irradiation or thermal processing, showing the thermal processing of HNCO-bearing ice to be the most favorable pathway. \cite{2004CP....302..259R} studied the HNCO:NH$_3$:H$_2$O=0.1:1:10 ice at 10~K via quantum modelling and laboratory methods. HNCO reacts with the formation of NH$_4^{+}$OCN$^{-}$ without irradiation or heating if enough H$_2$O molecules are present nearby. However, \cite{MD20} showed the formation of H$^{15}$NCO in H$_2$:CO:$^{15}$N$_2$ ices processed with 2 keV electron irradiation at 4~K; but neither $^{15}$NH$_3$ nor OC$^{15}$N$^-$ (the expected product of an acid–base reaction between H$^{15}$NCO and $^{15}$NH$_3$) was observed. \cite{MD24} extended the H$^{15}$NCO formation to higher temperatures. H$^{15}$NCO was detected in H$_2$:CO:$^{15}$N$_2$ ices deposited and irradiated at 4.7~K and 8~K, and in ices grown at 6~K and irradiated at temperatures up to 15~K. The H$^{15}$NCO yield decreases when the trapped H$_2$ abundance is low and at the highest irradiation temperature (15~K); H$^{15}$NCO was not detected in the ice both deposited and irradiated at 10~K.

\cite{2014ApJ...788...19J} carried out a set of experiments on UV irradiation and warm-up of ice analogues from 7~K that included H$_2$O, a carbon-bearing molecule (CO, CH$_3$OH, or CH$_4$) and a nitrogen source (NH$_3$ or N$_2$). Only the processing of H$_2$O:NH$_3$:CO=2:1:1 ice mixture led to a detectable amount of HNCO. \cite{2015MNRAS.446..439F} analyzed solid HNCO formation pathways in CO-rich ice analogues. They demonstrated that the yield of NH$_3$ versus HNCO depends on the flux of atomic hydrogen; a lower H flux favors the reaction of NH radicals with CO to form HNCO. \cite{2016MNRAS.460.4297F} reported efficient formation of HNCO in NO-bearing ices rich in CO, H$_2$CO and CH$_3$OH at 13 K upon simultaneous UV irradiation and hydrogenation under conditions simulating dark molecular clouds. \cite{2018MNRAS.475.1819F} showed that combined formation yields of HNCO and OCN$^{-}$ in CO- and CH$_3$OH-rich ices are significantly higher than in `H$_2$O-rich' ices. Measurements of HNCO/OCN$^{-}$ can also act as tracers for molecular nitrogen (N$_2$).

Most laboratory experiments focus on ice compositions that favor specific reactions, while significantly less attention is paid to the spectroscopic properties of HNCO in astrochemically relevant ices. Currently, band strengths and reference laboratory spectra are available only for pure HNCO and HNCO in H$_2$O environment \citep[see][ and references therein]{2025MNRAS.537.2918G}. 

Secure identifications, however, require extensive datasets of laboratory reference spectra for various molecular environments. Indeed, the 4.4--4.5 $\mu$m band was also explored by \cite{2024A&A...686A..71N,2026A&A...711A.240N}. The region was fit with only methyl cyanide (CH$_3$CN), ethyl cyanide (C$_2$H$_5$CN) and nitrous oxide (N$_2$O). The authors note HNCO as well as other CN-bearing molecules as potential candidates (vinyl cyanide C$_2$H$_3$CN, cyanogen C$_2$N$_2$ and HC$_n$N), but all those were dismissed based on their absorption profiles available from public databases \citep{2024A&A...686A..71N}. This points at a lack of reference spectra: HNCO is formed within the apolar environment in experiments with CO with a band centered at 2260 cm$^{-1}$ \citep{2018MNRAS.475.1819F}, while e.g., in IRAS20126 the peak position of 4.4--4.45 $\mu$m feature is at approximately 2259.5 cm$^{-1}$ \citep[Fig. 3,][]{2024A&A...686A..71N}.

We present transmission IR laboratory spectra of HNCO diluted in H$_2$O, CO, CO$_2$ and CH$_3$OH with 1:10 composition, which explore its spectral behavior in polar, apolar and methanol environments. We also present band strengths for HNCO in the mixtures. Further, we focus on identification of HNCO in JWST data near 4.42~$\mu$m ($\sim$2262 cm$^{-1}$) and constraining its molecular environment through fitting. We calculated the column density of HNCO using the relevant IR spectra for 12 apertures toward 11 protostars from the Mikulski Archive for Space Telescopes (MAST) database first explored by \cite{2026AA...706L...4K}. We also performed astrochemical modeling of HNCO and related species in the icy mantles of interstellar grains using the MONACO model described in \cite{Borshcheva_ea2025}.

\section{Methods} \label{sec:all_methods}
\subsection{Experimental} \label{sec:met:experimental}
All the transmission laboratory spectra used for solid HNCO identification were obtained using the Ice Spectroscopy Experimental Aggregate (ISEAge), a cryogenic ultra-high vacuum setup described in detail in \cite{2024ApJ...972L..10O}. In the main chamber a replaceable CaF$_2$ substrate is installed on a cold finger cooled by closed-cycle helium cryostat. With the Lakeshore 335 controller, we can control the substrate temperature in the range from 6.5~K to 305~K using resistive heating (50 $\Omega$). The transmission laboratory IR spectra were obtained with the Thermo Scientific Nicolet iS50 FTIR spectrometer operated within a 4000--840~cm$^{-1}$ (2.5--11.9~$\mu$m) range with 1~cm$^{-1}$ resolution. The individual deposition rates were controlled using the Stanford Research Systems RGA200 quadrupole mass spectrometer (QMS) via the calibration curves. The individual calibration curves for pure substances were obtained using FTIR and QMS following \cite{2023A&A...677A..13S}.

The HNCO ice analogues were grown on a CaF$_2$ substrate cooled down to 10 K, similarly to the protocol for binary CH$_4$-bearing mixtures from \cite{karteyeva2026infrared}. In this paper we present the transmission laboratory spectra of HNCO absorption in the various environments: HNCO:H$_2$O = 1:10, HNCO:CO = 1:10, HNCO:CO$_2$ = 1:10, HNCO:CH$_3$OH = 1:10. The~1:10 ratio makes HNCO a minor component in our mixtures, consistent with its low abundance constrained by \cite{2026AA...706L...4K}. The HNCO deposition rate was calibrated using the band strength of 129\texttimes10$^{-18}$~cm molecule$^{-1}$ at 10 K taken from \cite{2025MNRAS.537.2918G}. For H$_2$O, CO, CO$_2$ and CH$_3$OH we used the following band strengths: 2.2\texttimes10$^{-16}$ cm molecule$^{-1}$ \citep{2015MNRAS.451.2145B}, 9.07\texttimes10$^{-18}$ cm molecule$^{-1}$ \citep{2023MNRAS.522.3145G}, 1.18\texttimes10$^{-16}$ cm molecule$^{-1}$ \citep{2015ApJ...808L..40G} and 16.2\texttimes10$^{-18}$ cm molecule$^{-1}$ \citep{2024ApJ...970..108H}, respectively. Each co-deposition continued for 60 minutes and the HNCO deposition rate was fixed at 3\texttimes10$^{12}$~cm$^{-2}$~s$^{-1}$, resulting in total HNCO column density of 1.1\texttimes10$^{16}$~cm$^{-2}$.  After deposition the ice mixtures were warmed up at a rate of 1 K per minute. During the experiments the IR laboratory spectra were recorded every 45 seconds (averaging over 32 scans).

The HNCO monomer was synthesized from the thermal decomposition of the cyanuric acid (CNOH)$_3$ at 650 $^\circ$C under primary vacuum, following \cite{2013A&A...555A..13M}. Other compounds used are as follows: (CNOH)$_3$ (98$\%$, Sigma-Aldrich), CO (99.9999 $\%$, UgraPGS), CO$_2$ (99.9999 $\%$, UgraPGS), deionized H$_2$O, CH$_3$OH ($\geqslant$99.8~\%, Vekton). Gaseous CO and CO$_2$ were introduced directly into the dosing line from the commercially acquired gas bottles. H$_2$O and CH$_3$OH were purified by a freeze-pump-thaw cycle three times before each experiment.

\subsection{Observations and fitting procedure} \label{sec:met:obsfit}
We tried to confirm the presence of HNCO in interstellar ices using laboratory reference spectra and to constrain its molecular environment. The most prominent HNCO absorption band is located at 4.42~$\mu$m ($\sim$2262~cm$^{-1}$). In this work, we use the same set of fully processed open JWST NIRSpec spectra with identified N$_2$O and suggested HNCO features as presented in \cite{2026AA...706L...4K}. The observational data, anchor points and fitted local continua in 4.2--4.9 $\mu$m range for all investigated sources are shown in Fig. B.1. of \cite{2026AA...706L...4K}. Also shown is the red wing of $^{12}$CO$_2$ band (4.27 $\mu$m), $^{13}$CO$_2$ band (4.39 $\mu$m), OCN$^-$ band (4.6 $\mu$m) and CO band (4.67 $\mu$m). There is also a wide low-intensive H$_2$O libration mode ($\sim$3.7--5 $\mu$m) that overlaps with the 4.4--4.5 $\mu$m region. However, it blends with the signal and is accounted for by a conservatively selected continuum, as seen from Fig. B.1. in \cite{2026AA...706L...4K}. Several spectra were excluded from that sample due to low signal-to-noise ratio in the region of HNCO feature or strong contamination from gaseous H$_2$ emission in the 2270--2265 cm$^{-1}$ range. Therefore, the blue side of the fitted wavenumber range was limited by the H$_2$ emission for all sources, except of B1-b and Per-emb-27 (see Table \ref{app:a:shared_data} for the exact ranges). We also revisited the full 50 source sample \citep[MAST DOI: \href{https://doi.org/10.17909/tg2a-kv41}{10.17909/tg2a-kv41;}][]{2026AA...706L...4K} to search for sources with HNCO and no N$_2$O, but found none. The final sample of 12 spectra towards 10 Class 0 and 2 Class 0/I (B1-c, IRAS20126) protostars therefore represents the full subset of suitable data for HNCO analysis among the 50 protostars surveyed. All the sources are low-mass protostars \citep{Tobin2016, 2016ApJS..224....5F, Ortiz-Le2018}, except from IRAS20126, which is a massive protostar \citep{2016ApJ...823..125C}. For selected sources, \citet{2026AA...706L...4K} provide two extractions of the same ice feature from color-coded apertures: a spectrum from an emission-free aperture (labeled \textit{g}, green) and a spectrum in which gaseous CO lines were masked and replaced by interpolation (labeled \textit{w}, white). We use the \textit{g} spectra of HOPS-153 and HOPS-73 and both spectra of Per-emb-27, and keep this notation throughout the paper.

HNCO feature at 4.42~$\mu$m exhibits variations in shape and peak position depending on temperature and composition of the ice. To investigate the environment of HNCO we employed three variants according to the multi-layer ice model presented, e.g., by \cite{2015ARAA..53..541B}: (i) a polar H$_2$O-dominated layer, (ii) an apolar layer rich in CO, CO$_2$ and N$_2$, and (iii) a methanol-rich layer, where CH$_3$OH is efficiently formed after CO freeze-out at $A_v>9\pm3$ \citep[e.g.,][]{2011ApJ...729...92B}. Following the approach of \cite{2026AA...706L...4K}, we included laboratory spectra of N$_2$O in apolar environments (binary mixtures with CO, CO$_2$, N$_2$ and ternary mixtures with CO$_2$ and N$_2$ and CO$_2$ and CO) to fit the overlapping 4.45 $\mu$m N$_2$O band, which lies adjacent to the HNCO feature. The following N$_2$O spectra were used: N$_2$O:CO$_2$ = 1:20, N$_2$O:CO = 1:20, N$_2$O:N$_2$ = 1:20, N$_2$O:CO$_2$:CO = 1:15:5, and N$_2$O:CO$_2$:N$_2$ = 1:15:13.

In summary, the laboratory spectra are combined into four sets in the following way:
\begin{enumerate}
    \item Polar set: HNCO:H$_2$O=1:10 combined with spectra of N$_2$O in the apolar environments.
    \item Apolar set: HNCO:CO=1:10, HNCO:CO$_2$=1:10 combined with spectra of N$_2$O in the apolar environments.
    \item Methanol set: HNCO:CH$_3$OH=1:10 combined with spectra of N$_2$O in the apolar environments.
    \item Null set: N$_2$O spectra in apolar environments only, with no HNCO component. This set tests the null hypothesis that including HNCO does not improve the fit.
\end{enumerate}

HNCO- and N$_2$O-containing ice mixtures deposited at 10~K were heated up to 23~K. 18 spectra of each ice mixture were recorded during the warm-up, at 0.7~K temperature intervals. The lower limit of 10 K corresponds to the deposition temperature of the laboratory ices, while the upper limit is set to 23 K to ensure consistency between HNCO and N$_2$O laboratory spectra. Above 23~K N$_2$ can no longer be retained on the substrate on the timescale of the experiment. HNCO and N$_2$O spectra were grouped in combinations of equal temperature. For each combination of laboratory spectra at a given temperature $T_i$, the $\chi^2$ value was minimized between observational data and a linear combination of laboratory spectra:
\begin{equation}\label{chi2}
    \chi^2(T_i)=\sum^{N_{obs}}_{j=1}\frac{(\tau_{obs}^j-\tau_{lab}^j(T_i))^2}{\sigma^2_j},
\end{equation}
where $\tau_{obs}$ is the optical depth of the observational spectrum, and $\tau_{lab}(T_i)$ is the optical depth of a fitted linear combination of the laboratory spectra, $T_i$ is the temperature of the fitted combination, $N_{obs}$ is a number of points in the fitted segment of the observational spectrum, and $\sigma_j$ is the point-wise standard deviation of the observational spectrum. A sliding-window smoothed curve \citep[locally estimated weighted scatterplot smoothing,][]{Cleveland_1979} was subtracted from the original optical depth spectrum to estimate $\sigma_i$ from the residuals. The sliding window width was empirically chosen as 15 points for medium-resolution G395M spectra $(R \sim 1000)$ and 40 points for high-resolution G395H spectra $(R \sim 2700)$. These values are approximately the HWHM (half width at half maximum) of the 4.42 $\mu$m feature.

From the fits across the 10--23~K range, we selected the best-fit model for each source. Using the covariance matrix returned by the fitting routine \citep[\texttt{curve\_fit}, SciPy;][]{2007WRR....43.3423V,2020SciPy-NMeth}, we propagated uncertainties to derive the total HNCO and N$_2$O column densities as linear combinations of individual spectral components. We chose the corrected Akaike Information Criterion (AICc) for best fit model selection \citep{1974ITAC...19..716A,Sugiura01011978,10.1093/biomet/76.2.297}. Each set is characterized by its own AICc value:
\begin{equation}
    \mathrm{AICc}=\chi^{2}+2k+\frac{2k(k+1)}{n-k-1},
\end{equation}
where $k$ is the number of free parameters and $n$ is the amount of data points in the fitted section of the spectrum. Here the free parameters are fitted linear coefficients and offset $y_0$. The models are compared based on $\Delta_i$ values:
\begin{equation}
    \Delta_i = \text{AICc}_i - \text{AICc}_{min},
\end{equation}
where AICc$_{min}$ is the minimal AICc across the models. \cite{jabi} suggests the following level of empirical support of models based on $\Delta_i$ differences: $\Delta_i=$ 0--2: substantial, $\Delta_i=$ 4--7: considerably less, $\Delta_i>10$: essentially none. See Appendix \ref{app:A:fitstat} for reasoning behind selection of AICc over AIC or BIC (Bayesian Information Criterion) and goodness-of-fit tests. Additional statistical data, including fitted wavenumber range and $\sigma$ level by source, are given in Table \ref{app:a:shared_data}.

The relative support of each set is expressed as an Akaike weight
\begin{equation}
    w_i=\frac{\exp(-\Delta_i/2)}{\sum_j\exp(-\Delta_j/2)},
\end{equation}
where the summation is over $\Delta_i$ for all tested sets. The Akaike weight $w_i$ represents the probability that set $i$ is the best approximation among the tested sets, given the data \citep{jabi}. The weights for all tested sets for a single observational spectrum sum to unity, so they can be read directly as fractional support: a model with $w_i\simeq1$ is selected unambiguously, whereas comparable $w_i$ indicate that two or more ice environments remain statistically viable.

The temperature-dependent band strengths for column density estimations of HNCO in mixtures were derived from band area measurements based on the value from \cite{2025MNRAS.537.2918G} (see the band strengths in our Table~\ref{tab:hnco_bs}). The N$_2$O band strengths at 10~K are taken from \cite{2026AA...706L...4K} and converted to other temperatures.

\subsection{$^{12}$CO$_2$ column density estimation} \label{sec:res:corr:co2}
We used the $^{12}$CO$_2$ (15.2 $\mu$m) and $^{13}$CO$_2$ (4.39 $\mu$m) column densities from \cite{Brunken} to estimate $^{12}$CO$_2$ column densities from the $^{13}$CO$_2$ data presented by \cite{2026AA...706L...4K}. The Pearson's $\rho$ between $^{12}$CO$_2$ and $^{13}$CO$_2$ reveals a very strong correlation, with $\rho=0.993$ and a p-value below $10^{-6}$. Two-sided p-value was computed with a permutation test that randomly shuffles the pairings between the two variables ($10^7$ resamples). The column densities, however, are affected by band strength uncertainty, as noted by \cite{Brunken}. To align the data between the two papers, we use integrated optical depths (peak areas, $S$) to derive a linear formula that is free from band strength selection uncertainty. The peak areas scatterplot is presented in panel \textit{a} of Fig. \ref{fig:corr}. We report the following formula for peak areas: 
\begin{multline}\label{corr}
    S\left(^{12}\text{CO}_2, 15.2\, {\rm \mu m} \right)=(16.0\pm0.5)\cdot S\left(^{13}\text{CO}_2, 4.39\, {\rm \mu m} \right)-\\-(4.0\pm1.9)
\end{multline}
The $^{13}$CO$_2$ column densities from \cite{2026AA...706L...4K} were converted to peak areas using the band strength from their Table C.2, and the $^{12}$CO$_2$ peak areas were calculated using equation (\ref{corr}). Then, the $^{12}$CO$_2$ column densities were obtained using the same band strength of 1.6\texttimes$10^{-17}$ cm molecule$^{-1}$ as in \cite{Brunken}. We note that the derived $^{12}$CO$_2$ values should be treated as reliable approximations due to the high relative uncertainty of the free coefficient in equation (\ref{corr}). The $^{12}$CO$_2$/$^{13}$CO$_2$ ratio derived from $^{13}$CO$_2$ column densities of \cite{2026AA...706L...4K} using the $^{12}$CO$_2$ and $^{13}$CO$_2$ band strength from \cite{Brunken} are in range of 86--117 with a median of 108. This range aligns with the $^{12}$CO$_2$/$^{13}$CO$_2$ ratio range of 70--125 with a median of 100 reported in \cite{Brunken}. 

\section{Results} \label{sec:results:all}
\subsection{Laboratory} \label{sec:res:lab}
\begin{figure}[ht]
   \centering
    \includegraphics[width=\hsize]{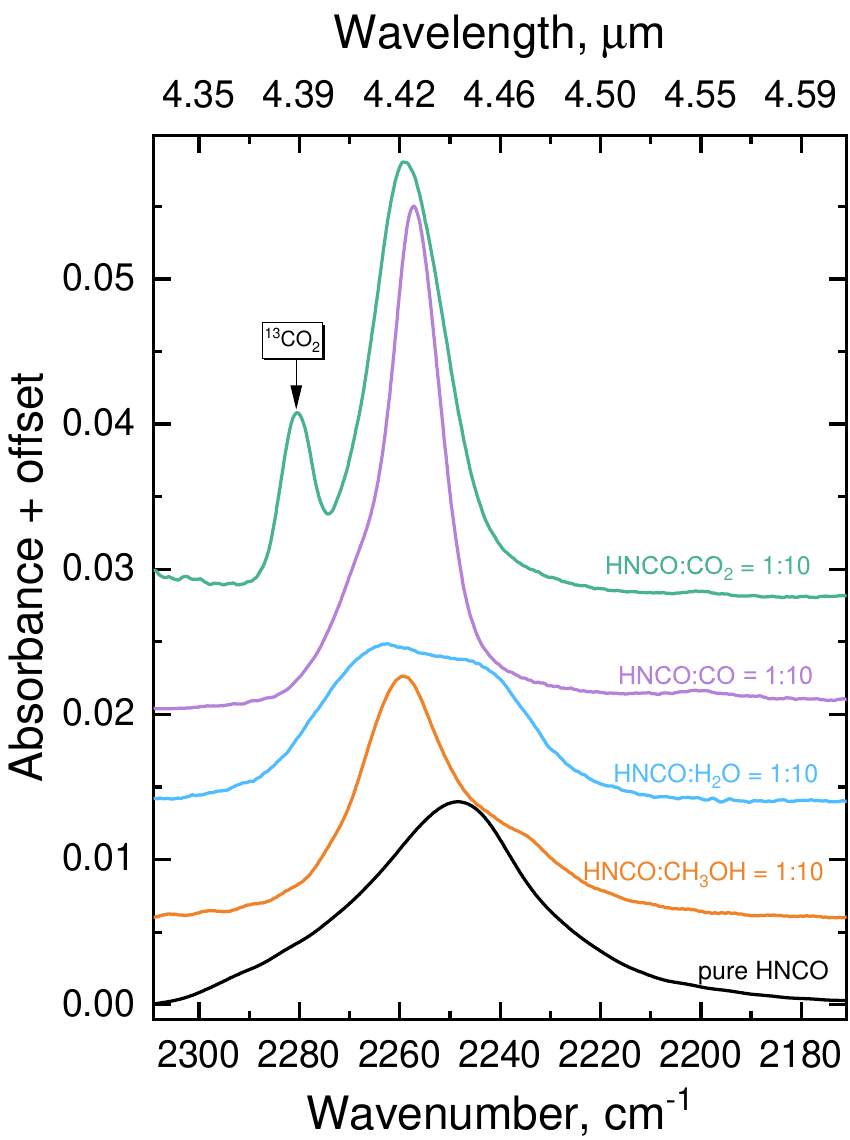}
    \caption{Position and profile of the $\nu_2$ absorption feature of HNCO in various mixtures at 10 K: black shows pure HNCO, orange the HNCO:CH$_3$OH, cyan the HNCO:H$_2$O, purple the HNCO:CO, green the HNCO:CO$_2$. The column density of HNCO in all the mixtures equals to 1.1\texttimes10$^{16}$~cm$^{-2}$.} 
    \label{fig:labspec}
\end{figure}

The laboratory IR transmission spectra of the HNCO $\nu_2$ mode for 1:10 mixtures with H$_2$O, CO, CO$_2$, and CH$_3$OH at 10~K are presented in Fig. \ref{fig:labspec}. The peak position of HNCO in mixtures is blueshifted relative to pure HNCO: by 8 cm$^{-1}$ for both CO and CO$_2$, by 10 cm$^{-1}$ for CH$_3$OH, and by 13 cm$^{-1}$ for H$_2$O matrix. The NCO asymmetric stretching mode ($\nu_2$) feature of HNCO in CO and CO$_2$ is a single asymmetrical peak. In CH$_3$OH, a red shoulder develops, and in H$_2$O the feature consists of two unresolved components. After the depositions ices were heated at a rate of 1 K per minute and additional spectra were taken and used for the fit. 

Over 10--23~K only the HNCO:CO profile changes: from 15 K to 23 K the band strength of 4.42 $\mu$m feature gradually decreases by 10\%. This is likely caused by crystallization of the CO matrix. The 3450--3000 cm$^{-1}$ (2.89--3.33 $\mu$m) range is assigned to the NH stretch mode \citep[$\nu_1$, ][]{1997JChPh.107.9764B}. Water and methanol peaks near 3.3 $\mu$m overlap with NH mode of HNCO (see Appendix~ \ref{app:B:tablesandplots}). The intensity of $\nu_1$ peak is less than that of $\nu_2$ peak by approximately a factor of 5. We also derived band strengths for HNCO $\nu_2$ band in various environments at 10 K, see Table~\ref{tab:hnco_bs}. The band strength values were determined by scaling the band strength value for pure HNCO from \cite{2025MNRAS.537.2918G}, under the assumption that the column density is the same in all samples. The method is described in more detail in \cite{2022AA...665A..89R} and \cite{2023A&A...677A..13S}. Our band strength for HNCO:H$_2$O=1:10 at 10 K agrees with the 97.9\texttimes10$^{-18}$~cm molecule$^{-1}$ value reported by \cite{2025MNRAS.537.2918G}. 

\begin{deluxetable}{ccc}[t]
\tablecaption{HNCO $\nu_2$ absorption band strength values derived in this work and used for the fit.\label{tab:hnco_bs}}
\tablewidth{0pt}
\tablehead{
\colhead{Ice} & \colhead{Peak Position} & \colhead{Band Strength} \\
composition        & (cm$^{-1}$)  & (cm molecule$^{-1}$)          }
\startdata
HNCO:CO         & 2257      & 11.3\texttimes10$^{-17}$ \\
HNCO:CO$_2$     & 2259      & 12.1\texttimes10$^{-17}$ \\
HNCO:CH$_3$OH   & 2259      & 11.2\texttimes10$^{-17}$ \\
HNCO:H$_2$O     & 2262      & 9.7\texttimes10$^{-17}$  \\
\enddata
\tablecomments{All mixed ices are presented in HNCO:X=1:10 ratio at 10~K. The estimated errors of band strengths are $\le$7\%.} 
\end{deluxetable}

\subsection{Fitting}\label{sec:res:fit}

\begin{figure*}[ht]
    \centering
    \includegraphics[width=0.99\textwidth]{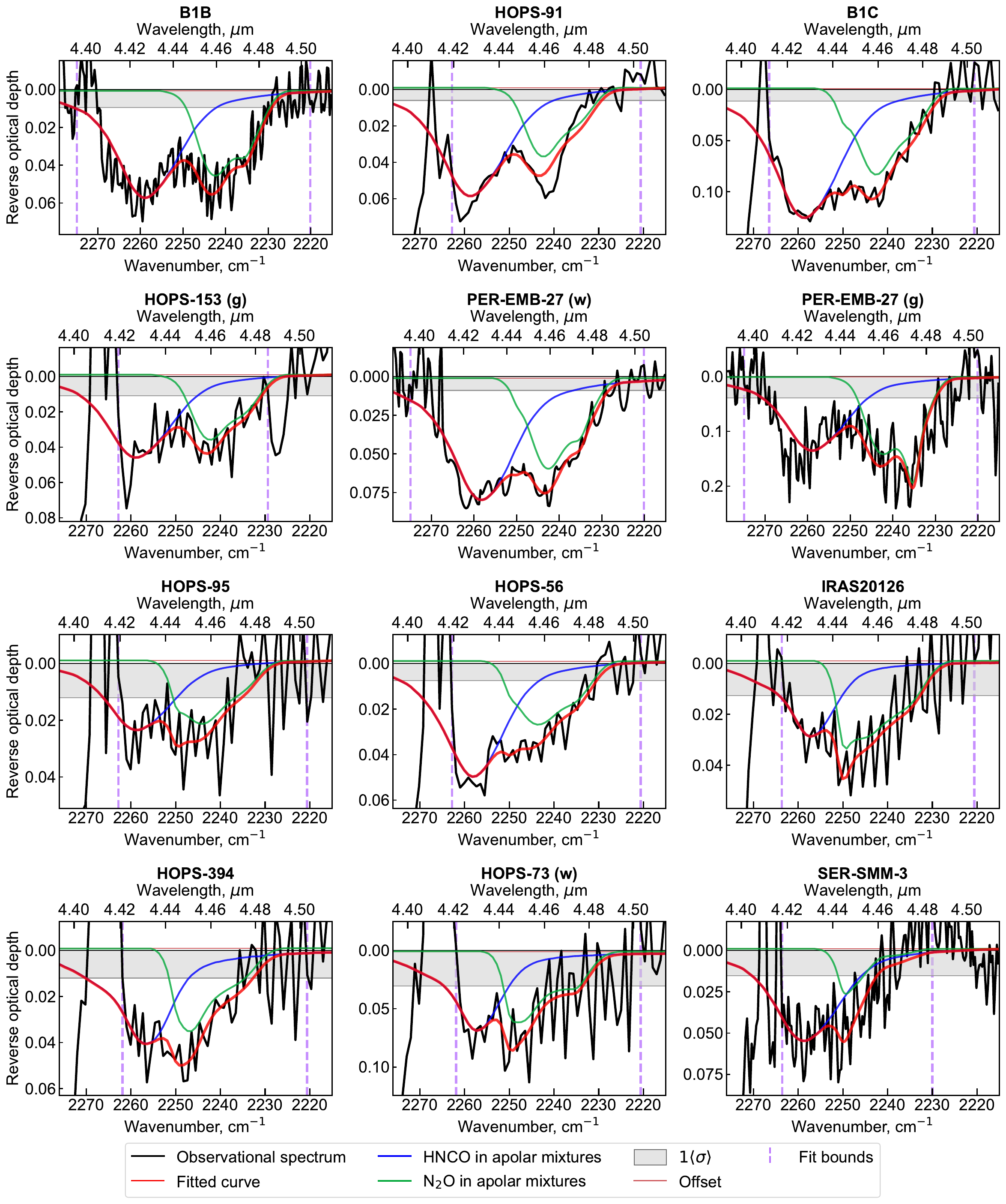}
    \caption{Observation data in 2270--2215 cm$^{-1}$ (4.4--4.55 $\mu$m) region and the best-fit set of spectra. Blue --- combined HNCO feature of the Apolar set, green --- combined N$_2$O feature. The dashed vertical lines indicate the bounds of the fitted range. The letters \textit{g} and \textit{w} denote the colors of the apertures from  \cite{2026AA...706L...4K}, also see Section \ref{sec:met:obsfit}.}
    \label{fig:main_fit}
\end{figure*}

We fit the 4.4--4.5 $\mu$m feature in 12 spectra first presented in \cite{2026AA...706L...4K} using relevant HNCO and N$_2$O mixtures. We confirm HNCO detections with laboratory spectra in all 12 spectra; the fits of Apolar set are shown in Fig. \ref{fig:main_fit}, derived column densities, abundances relative to solid CO and approximate abundances relative to solid H$_2$O are listed in Table \ref{tab:coldens}. The 3.3 $\mu$m OH stretch feature is saturated on the spectra and usually is also distorted by grain size effects \citep{2024NatAs...8..359D}, 4.4 $\mu$m OH libration is a wide low intensity feature strongly affected by continuum selection. Other H$_2$O features belong to the MIRI-MRS range and require complex additional analysis, which is out of the scope of this paper \citep[see e.g.,][]{Petrashkevich_ea2026}. Therefore, to estimate HNCO abundance to the order of magnitude and provide a common reference value we calculated the approximate abundances by converting the X/$^{12}$CO$_2$ ratio to X/H$_2$O using the median H$_2$O/$^{12}$CO$_2$ ratio of 0.28 \citep{2015ARAA..53..541B}. See Section \ref{sec:res:corr:co2} for derivation of $^{12}$CO$_2$ column density. We report the approximate abundances for HNCO and N$_2$O of 0.04--0.15\% and 0.04--0.19\%, respectively. These estimates fall within the upper limits derived by \cite{2004A&A...415..425V}. Note that \cite{2010ApJ...725.2101Q} modeled the abundance of HNCO with respect to $n_{H_2}$ in the gas and on the dust particles (see their Fig. 5). The HNCO gas/ice ratio is approximately 3\texttimes10$^{-4}$ at the age of 10$^{5}$ years. Gaseous HNCO column density is $\sim2.4\times10^{13}$ cm$^{-2}$ towards B1-b \citep{2009A&A...494L..13O}. Thus, the approximate value for solid HNCO towards B1-b is $\sim8\times10^{16}$ cm$^{-2}$, of the same order of magnitude with our estimates.

The $\Delta_i$ values were calculated for the spectral sets (see Section \ref{sec:disc:met_fit}). The $\nu_2$ feature varies sufficiently between the laboratory spectra of different mixtures to determine the prevailing HNCO environment. At the same time, the spectra of laboratory mixtures vary only slightly in the considered temperature range of 10--23 K, and this prevents us from drawing any conclusions from the fitted temperatures. The fits with different sets are shown in Fig. \ref{fig:app_fit}. Based on $\Delta_i$ values, seven of the twelve spectra show strong evidence towards HNCO in the apolar ice layer $(\Delta_i>10)$. For B1-b and HOPS-394 there is considerably less evidence towards HNCO in the methanol environment than for HNCO in the apolar ice layer ($\Delta_i=4.2$ and $\Delta_i=3.4$, respectively). The spectra for HOPS-153 and HOPS-73 are fit equally well with the Apolar and Methanol sets ($\Delta_i<0.5$), and there is less evidence for the Water set in these two cases. Finally, HOPS-95 is fit equally well by all three spectral sets ($\Delta_i\le1$). The Null model, which does not contain any HNCO spectra is not supported based on $\Delta_i\sim 120$--4664, therefore we reject the null hypothesis. More statistical data, including the best-model AICc values, can be found in Appendix \ref{app:A:fitstat}. 

The Polar set consistently fails to fit the data: the 4.42~$\mu$m HNCO feature in H$_2$O is too broad and not resolved enough to match the observational data. The Methanol set provides a reasonable alternative to the Apolar set in a few cases. We conclude that the Apolar set is the most consistent with the observational data based on the analysis of our sample. We therefore adopt the fit results for HNCO in the apolar layer for further analysis as a prevailing HNCO environment.

\begin{deluxetable*}{ccccccccc}[htbp]
\tablecaption{Column densities and model comparison for fits with HNCO and N$_2$O.\label{tab:coldens}}
\tablewidth{0pt}
\tablehead{
\colhead{Object} & \colhead{\textit{N}(HNCO)}&\colhead{ $\frac{N(\text{HNCO})}{N(\text{CO})}$} &\colhead{ $\frac{N(\text{HNCO})}{N(\text{H}_2\text{O})}$} & \colhead{\textit{N}(N$_2$O)} &\colhead{ $\frac{N(\text{N}_2\text{O})}{N(\text{H}_2\text{O})}$ }&\multicolumn{3}{c}{$\Delta_i={\rm AICc}_i-{\rm AICc}_{min}$} \\
\colhead{} & \colhead{$10^{16}$ cm$^{-2}$}& \colhead{\%} & \colhead{\%} & \colhead{$10^{16}$ cm$^{-2}$}& \colhead{\%} & \colhead{Apolar}& \colhead{Methanol} & \colhead{Water}} 
\startdata
B1-b           & 0.99 $\pm$ 0.04 &0.21&0.09& 0.94 $\pm$ 0.06 &0.08& 0.0 & 4.2 & 125.9 \\ 
HOPS-91        & 1.05 $\pm$ 0.12&0.27 &0.07& 0.71 $\pm$ 0.06 &0.04& 0.0 & 93.2 & 113.5 \\ 
B1-c           & 2.28 $\pm$ 0.19 &0.59&0.09& 1.82 $\pm$ 0.15 &0.07& 0.0 & 260.1 & 721.2 \\
HOPS-153 (g)   & 0.82 $\pm$ 0.14 &0.28&0.04& 0.72 $\pm$ 0.19 &0.04& 0.1 & 0.0 & 6.0 \\ 
Per-emb-27 (w) & 1.48 $\pm$ 0.09&0.50 &0.08& 1.27 $\pm$ 0.04 &0.07& 0.0 & 59.2 & 478.4 \\ 
Per-emb-27 (g) & 2.34 $\pm$ 0.19 &0.90&0.11& 3.58 $\pm$ 0.25 &0.17& 0.0 & 46.0 & 147.1 \\ 
HOPS-95        & 0.43 $\pm$ 0.24 &0.24&0.05& 0.56 $\pm$ 0.15 &0.06& 0.9 & 0.0 & 1.0 \\ 
HOPS-56        & 0.87 $\pm$ 0.07&0.54 &0.09& 0.71 $\pm$ 0.10 &0.08& 0.0 & 14.8 & 24.2 \\ 
IRAS20126      & 0.47 $\pm$ 0.11& 0.36&0.06& 0.91 $\pm$ 0.17 &0.11& 0.0 & 10.4 & 12.9 \\ 
HOPS-394       & 0.80 $\pm$ 0.15 &0.63&0.15& 0.85 $\pm$ 0.16 &0.16& 0.0 & 3.4 & 5.8 \\ 
HOPS-73 (w)    & 1.04 $\pm$ 0.28& 1.10&0.13& 1.56 $\pm$ 0.43 &0.19& 0.0 & 0.4 & 3.0 \\ 
Ser-SMM-3      & 0.98 $\pm$ 0.25& 1.53&0.15& 0.44 $\pm$ 0.20 &0.07& 0.0 & 56.8 & 115.0 \\ 
\enddata
\tablecomments{Column densities are given for the apolar ice model. $\Delta_i=0$ corresponds to the best HNCO ice fitting set (see Section~\ref{sec:met:obsfit}, the best absolute AICc values are listed in Table \ref{app:a:specific_data}). The apertures and designations are taken from \protect\cite{2026AA...706L...4K}. Approximate abundances relative to solid H$_2$O are derived from abundances relative to $^{12}$CO$_2$ using a scale coefficient of 0.28 for CO$_2$/H$_2$O in low mass young stellar objects \protect\citep{2015ARAA..53..541B}.}
\end{deluxetable*}

\subsection{Correlation analysis} \label{sec:res:corr:corr}
\begin{figure*}[ht]
   \centering
    \includegraphics[width=\hsize]{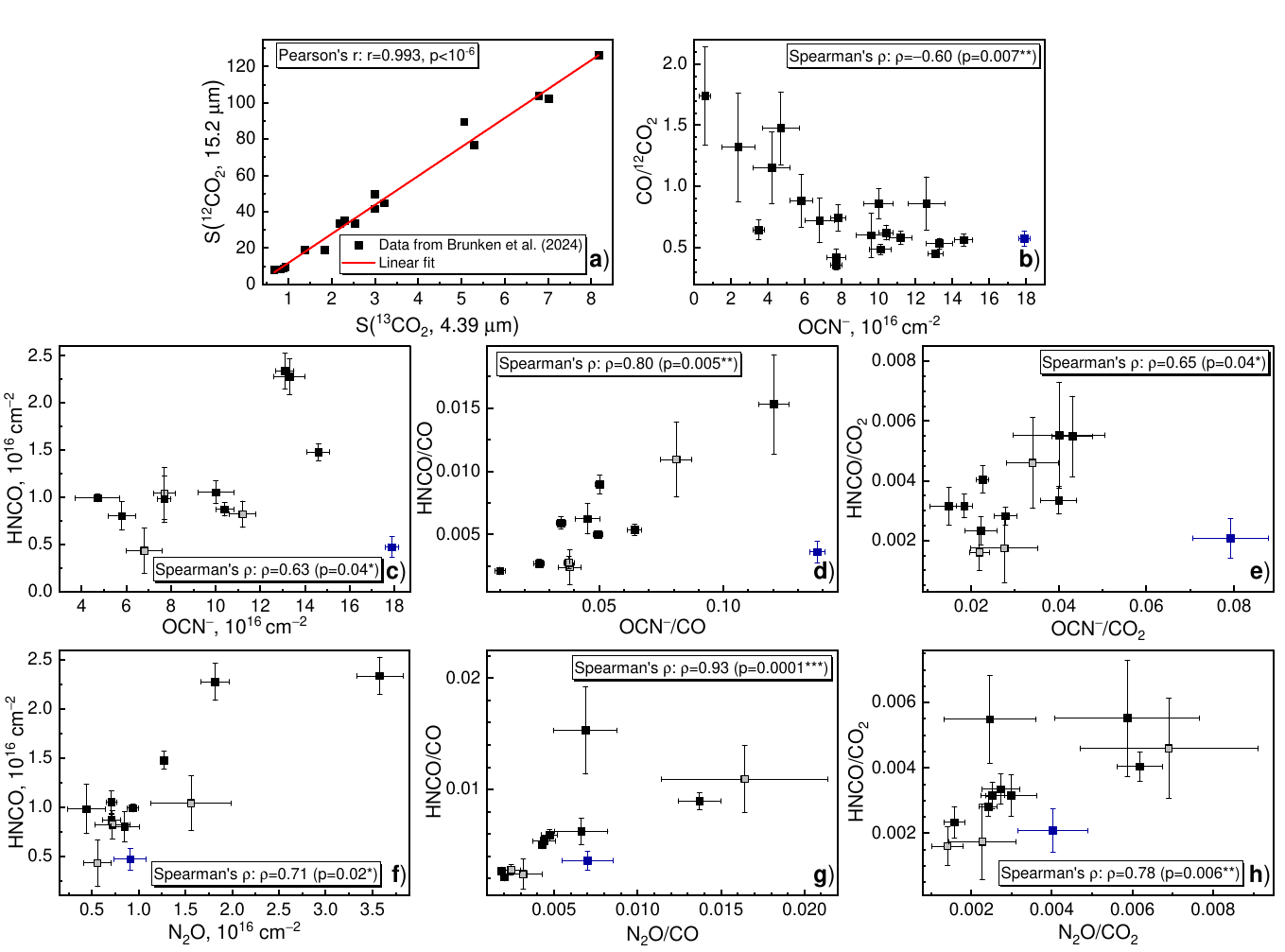}
    \caption{Correlations and scatterplots. The blue point, massive protostar IRAS20126, is excluded from the correlation analysis of the low-mass protostar sample, see Section \ref{sec:res:corr:corr}. Asterisks mark the significance: one asterisk indicates $p<0.05$, two indicate $p<0.01$, three indicate $p<0.001$.} 
    \label{fig:corr}
\end{figure*}

To further constrain the molecular environment and formation history of HNCO, the correlation coefficients are derived between column densities of HNCO, OCN$^-$ and N$_2$O. We used the sample of 20 CO, CO$_2$ and OCN$^-$ column density points from \cite{2026AA...706L...4K} and a sample of 12 HNCO and N$_2$O column density points derived in this work (Table \ref{tab:coldens}). Note that the Apolar set column densities are used for HOPS-153 and HOPS-95 for consistency, the preference against the Methanol set was given due to $\Delta_i=\text{AICc}_\text{Apolar}-\text{AICc}_\text{Methanol}<2$ (see Section \ref{sec:met:obsfit}). $^{12}$CO$_2$ column densities were calculated based on $^{13}$CO$_2$ column densities from \cite{2026AA...706L...4K}, see panel \textit{a} of Fig. \ref{fig:corr}, using the formula we derived from $^{12}$CO$_2$ and $^{13}$CO$_2$ data presented in \cite{Brunken}, see Section \ref{sec:res:corr:co2}.  

For correlation analysis we consider only the low-mass protostars, excluding IRAS20126. IRAS20126 is a massive protostar \citep{2016ApJ...823..125C}, likely more affected by thermal processing than the rest of the sample \citep{2015ARAA..53..541B}. This is also consistent with its OCN$^-$ abundances relative to low-mass sources. Thermal processing is beneficial to production of OCN$^-$ \citep[e.g., ][]{2004A&A...415..425V}, and both OCN$^-$/CO and OCN$^-$/CO$_2$ ratios for IRAS20126 exceed the median values for the low-mass sample of 11 by factors of 3 and 2.9, respectively. For a low-mass sample the OCN$^-$/CO ratio is $0.05\pm0.03$ (median $=0.05$); the OCN$^-$/CO$_2$ ratio is $0.03\pm0.01$ (median $=0.03$), where the uncertainties are the sample standard deviation. 

The three sources from \cite{2026AA...706L...4K}, namely B335, EDJ2009-183, and 2MASSJ17112317, in which HNCO feature was not found have the lowest OCN$^-$ abundance across the sample of 19 low-mass protostars and IRAS20126, which indicates low processing of the ice. This is also supported by a negative Spearman correlation between CO/$^{12}$CO$_2$ and OCN$^-$ with $\rho=-0.60$ (19 points, $p=0.007$\footnote{The amount of data points is relatively low and for all correlations two-sided p-values were computed with a permutation test that randomly shuffles the pairings between the two variables ($10^7$ resamples).}). The CO/$^{12}$CO$_2$ ratio traces the balance between the apolar and polar ice layers and OCN$^-$ content increases as CO oxidizes to $^{12}$CO$_2$, lowering the ratio. The corresponding scatterplot is presented in panel $b$ of Fig. \ref{fig:corr}. 

We report HNCO/OCN$^-$ ratio in the range of 0.06--0.21 (0.03 for IRAS20126) based on HNCO column densities derived in this work and OCN$^-$ data from \cite{2026AA...706L...4K}. There is also a positive Spearman correlation with $\rho=0.63$ (11 points, $p=0.04$) between HNCO and OCN$^-$ if the highest abundance OCN$^-$ source, IRAS20126 (see panel $c$ in Fig. \ref{fig:corr}). The correlation is stronger if the column densities are normalized to CO ($\rho=0.80$, $p=0.005$), and stays comparable if normalized to $^{12}$CO$_2$ ($\rho=0.65$, $p=0.04$) (see panels $d,e$ in Fig.~\ref{fig:corr}). \cite{2026AA...706L...4K} reported detection of N$_2$O and nondetection of HNCO in B335, EDJ2009‑183, and 2MASSJ17112317. We note that these sources and B1‑b have the lowest OCN$^-$ abundance across their sample. However, we detected both HNCO and N$_2$O in B1‑b. In the 19‑source sample, B1‑b has the highest CO and median CO$_2$ column density, indicating a more dominant apolar layer compared to the other three low‑OCN$^-$ sources.

In our previous work we analyzed a sample of 50 sources \citep[MAST DOI: \href{https://doi.org/10.17909/tg2a-kv41}{10.17909/tg2a-kv41;}][]{2026AA...706L...4K} and found N$_2$O to be accompanied by HNCO in 13 of 16 sources in which N$_2$O was detected. We found that HNCO and N$_2$O are positively correlated with $\rho=0.71$ (11 points, $p=0.02$). The correlation is stronger for HNCO/CO and N$_2$O/CO $(\rho=0.93,\ p=0.0001)$ than for HNCO/$^{12}$CO$_2$ and N$_2$O/$^{12}$CO$_2$ $(\rho=0.78,\ p=0.006)$. The scatterplots are presented in panels $f$-$h$ of Fig. \ref{fig:corr}.

Overall, the correlation analysis points towards the CO-related origin of OCN$^-$, HNCO and N$_2$O, and highlights a close relation between HNCO and N$_2$O. The OCN$^-$ result is in line with the established origin of OCN$^-$ from CO that resides in a polar layer of the ice \citep{2004A&A...415..425V,2022ApJ...941...32B}. 

\subsection{Modeling}\label{sec:res:model}
The composition of icy mantles of interstellar grains is sensitive to the history of physical conditions during the evolution of a protostellar cloud~\citep[see e.g.][]{Clement_ea23}. Since in this work we are interested in accurate reproduction of the layered structure of icy mantles, we employ the non-diffusive rate equations-based model MONACO described in~\citet[][]{Borshcheva_ea2025}, coupled with the time-dependent evolution of physical conditions in a 0D parcel of gas and dust as described in~\citet[][]{JimenezSerra_ea25}. The modeling was performed in two stages. The first is the translucent-cloud stage, as in \cite{Borshcheva_ea2025}, and the second is a one-dimensional gravitational collapse of a spherically symmetric protostellar core, during which the density and radial position of a gas parcel evolve with time.  

During the collapse stage, which lasts for $10^6$~yr, the density changes according to \cite{Rawlings_ea1992}. The visual extinction is related to the hydrogen column density by $A_V = N_H /(1.6\times10^{21}\ {\rm cm^{-2}})$, and the temperature as a function of $A_V$ is calculated according to \cite{Garrod_Pauly2011} in dense regions and according to \cite{Hocuk_ea2017} in less dense regions where $A_V \lesssim 8.9$. The final radial number density profile at the end of this stage is $n({\rm H_2}) = 10^9\ {\rm cm^{-3}}\,(r/{\rm AU})^{-3/2}$ \citep{Petrashkevich_ea2026} between 0.5 AU and 8000 AU. The latter value is a typical outer radius of a protostellar envelope \citep[e.g.][]{Jorgensen_ea2009}.

The chemical network was updated to include new reactions involving $\rm HNCO$ and $\rm N_2O$, including the barrierless reaction $\rm HNCO + NH_3 \rightarrow NH_4^+OCN^-$ and a more detailed treatment of the reaction $\rm CO + OH$ following \cite{Molpeceres_ea2025}. Previously, this reaction was used in our network as a simplified representation of the two-step process $\rm CO + OH \rightarrow HOCO,\ HOCO + H \rightarrow CO_2 + H_2$ forming $\rm CO_2$ ice. However, according to \cite{Molpeceres_ea2025}, hydrogenation of $\rm HOCO$ preferentially produces $\rm CO + H_2O$ and $\rm HCOOH$ rather than $\rm CO_2$. Formation of $\rm CO_2$ is primarily associated with the \textit{c-}$\rm HOCO + H$ channel, whereas \textit{t-}$\rm HOCO$ is approximately 4 kcal/mol more stable than \textit{c-}$\rm HOCO$. Consequently, the population is expected to be dominated by the \textit{t-} isomer. Since our network does not distinguish between the two isomers, we adopt an approximate branching ratio of $\rm CO + H_2O : HCOOH : CO_2 + H_2 = 5 : 4 : 1$.

The major source of $\rm HNCO$ ice in our model is the barrierless surface reaction $\rm H + OCN \rightarrow HNCO$. In turn, the $\rm OCN$ radical is produced by the surface reaction $\rm N + HOCO \rightarrow OH + OCN$, where $\rm HOCO$ is formed in the reaction of carbon monoxide with the hydroxyl radical. HNCO is primarily destroyed by the reaction $\rm HNCO + H \rightarrow NH_2CO$ \citep{Song_Kaestner2016} and by photolysis. The reaction $\rm HNCO +NH_3 \rightarrow NH_4^+OCN^-$, which is treated as barrierless, is slower than the hydrogenation of HNCO, as the latter can proceed through quantum tunneling. The reaction $\rm NH + CO \rightarrow HNCO$ is inefficient in our model because of its large activation barrier of 2870~K \citep{Jin_Garrod2020}. The resulting solid $\rm HNCO$ abundance relative to solid $\rm H_2O$ is in the range of 0.06--0.11\% throughout the core radial profile, in agreement with our fitting results. For the primary $\rm HNCO$ destruction reaction, $\rm HNCO + H \rightarrow NH_2CO$, \cite{Song_Kaestner2016} report an activation barrier of 31.8--36.5 kJ\,mol$^{-1}$ for the Eley--Rideal mechanism and 37.9 kJ\,mol$^{-1}$ for the Langmuir--Hinshelwood mechanism. In our network, we adopt a barrier of 4000~K (33.3~kJ\,mol$^{-1}$).

Atomic nitrogen ($\rm N$) plays an important role in the formation of both $\rm HNCO$ and $\rm N_2O$, and its primary source is freeze-out from the gas phase. Solid $\rm N_2O$ in our model forms via the reaction $\rm N + NO \rightarrow N_2O$ \citep{Jin_Garrod2020}. Its abundance relative to solid $\rm H_2O$ is 0.01--0.03\%. It lies near the lower end of the range obtained from the fitting results, likely due to the limited N$_2$O-related chemical network adopted. Other possible formation pathways for $\rm N_2O$ exist that are not included in our model. For example, the barrierless reaction $\rm N_2 + O(^1D)$ \citep{Jamieson_ea2005} involving excited atomic oxygen may also contribute to $\rm N_2O$ production. Including such pathways could further improve the agreement between the modeled and observed abundances of solid $\rm N_2O$.

\begin{figure}[ht]
  \includegraphics[width=\linewidth]{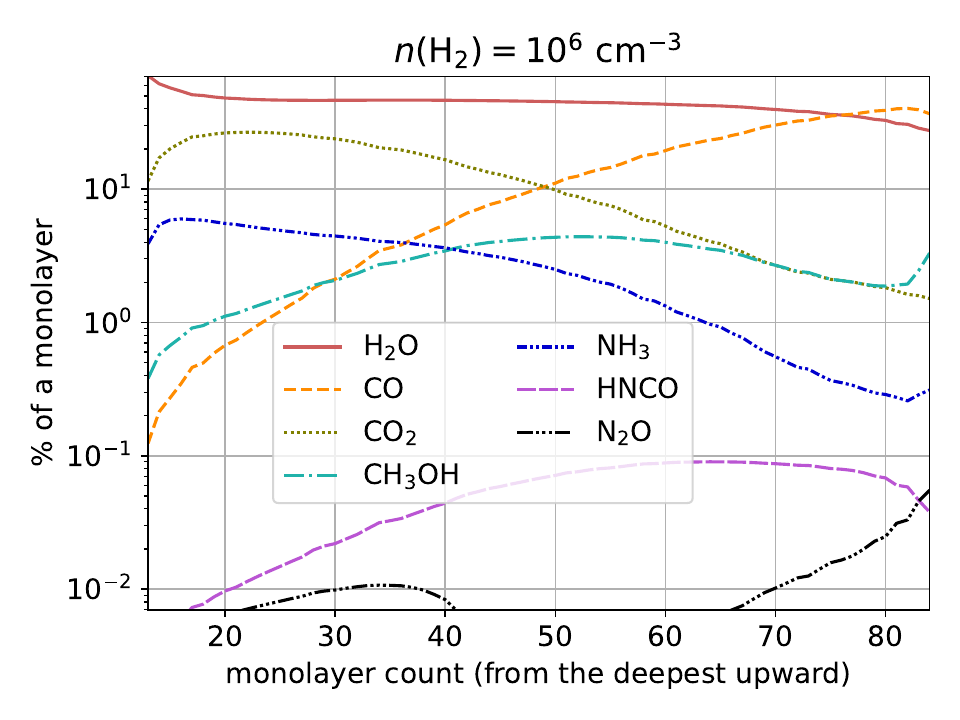}
\caption{Layer-by-layer ice structure in the gravitational collapse model of a protostellar core for the location with final $\rm H_2$ number density of $10^6\ \mathrm{cm^{-3}}$. For clarity, only selected species are displayed.
} \label{fig:model:lbl}
\end{figure}

In Figure \ref{fig:model:lbl} we show the layer-by-layer ice structure at a location with the final modeled H$_2$ number density of 10$^6$ (100 au from the core center). The model reproduces the increase in the CO abundance relative to H$_2$O toward the outer layers, which resembles the layered ice structure \citep{2015ARAA..53..541B}. At the end of the collapse stage, the gas density increases rapidly, while the gas and dust temperatures decrease, causing catastrophic CO freeze-out. Also, less CO$_2$ is present in the outer layers than in the inner ones, qualitatively agreeing with the observational distribution of CO$_2$ between polar and apolar ice layers derived in \cite{2008ApJ...678.1005P}. The HNCO content increases with the increase of CO, except for the outermost layers, where CO abundance approaches or exceeds that of $\rm H_2O$. In addition, CH$_3$OH and HNCO exhibit similar behavior, consistent with both species being products of CO chemistry. Ammonia (NH$_3$) abundance decreases toward the outer layers; however, the connection between the behaviors of NH$_3$ and HNCO is not straightforward, as the reaction $\rm HNCO + NH_3 \rightarrow NH_4^+OCN^-$ is not the dominant pathway for HNCO destruction in our model.

\section{Discussion} \label{sec:disc}
\subsection{HNCO vs CH$_3$CN} \label{sec:disc:met_fit}
Based on the analysis of the absorption in the 4.4--4.52 $\mu$m range \cite{2024A&A...686A..71N} reported first tentative detections of CH$_3$CN, C$_2$H$_5$CN and N$_2$O toward HOPS 153, HOPS 370 and IRAS20126, and upper limits toward IRAS 16253 and B335. The estimated column densities and upper limits for these sources are in range of (20--43)\texttimes10$^{16}$ cm$^{-2}$ for CH$_3$CN and (7.1--14)\texttimes10$^{16}$~cm$^{-2}$ for C$_2$H$_5$CN. No estimates for N$_2$O were provided. \cite{2026A&A...711A.240N} reported column densities in (24--140)\texttimes10$^{16}$ cm$^{-2}$ range for CH$_3$CN and (8.9--48)\texttimes10$^{16}$ cm$^{-2}$ for C$_2$H$_5$CN toward B1-b, B1-c and IRAS 2A (=Per-emb-27).

In this work we fit the B1-b, Per-emb-27 (w) and Per-emb-27 (g) spectra with an extended dataset that includes both new laboratory spectra of HNCO and N$_2$O and a spectrum of CH$_3$CN:H$_2$O=1:10 mixture at 15~K from \cite{2022AA...665A..89R}. Fits with CH$_3$CN were performed only on these three observational spectra, as the H$_2$ emission feature in the $\sim$2270--2265 cm$^{-1}$ range overlaps with CH$_3$CN feature in spectra of other sources. The CH$_3$CN:H$_2$O mixture extends our laboratory dataset on the blue side with the CN stretching $\nu_2$ feature of CH$_3$CN at 4.41 $\mu$m. Moreover, this is the only band strength for CH$_3$CN in mixture listed in \cite{2022AA...665A..89R}, $A(\text{CH}_3\text{CN})=3.325\times10^{-18}$ cm molecule$^{-1}$. The only available spectrum in LIDA (Leiden Ice Database for Astrochemistry\footnote{\url{http://icedb.strw.leidenuniv.nl/}}) from \cite{2022AA...665A..89R} is of pure CH$_3$CN, thus we adopted the CH$_3$CN:H$_2$O spectrum from their Fig. 3. For Null set with CH$_3$CN only HNCO spectra were excluded. We used the same fitting procedure as described in Section~\ref{sec:met:obsfit}.

\begin{figure*}
    \centering
    \includegraphics[width=\textwidth]{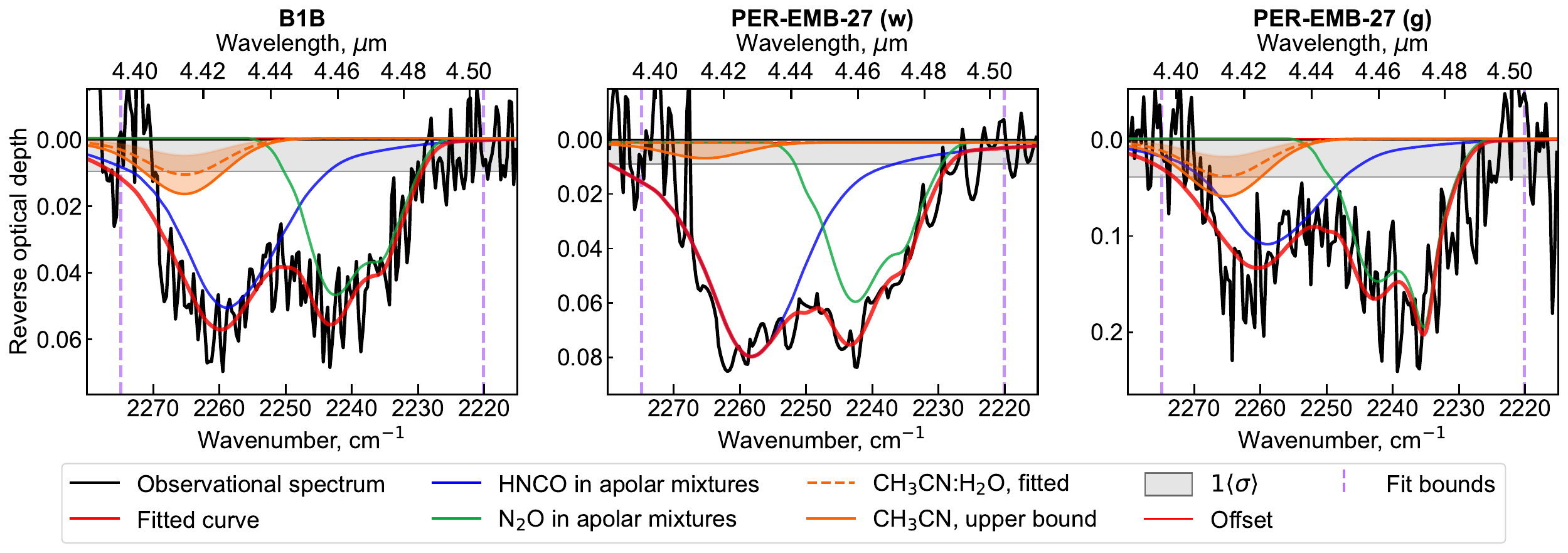}\label{CH3CN_fit}
    \caption{Observation data in 2280--2215 cm$^{-1}$ (4.4--4.55 $\mu$m) region and the best-fit set of spectra including CH$_3$CN. Blue --- combined HNCO feature of the Apolar set, green --- combined N$_2$O feature. The dashed orange line corresponds to the fitted column density of CH$_3$CN, the filled orange sector --- to the CH$_3$CN fit uncertainty and the solid orange line is the upper bound for CH$_3$CN (see Section \ref{sec:disc:met_fit}). The dashed vertical lines indicate the bounds of the fitted range. The letters \textit{g} and \textit{w} denote the colors of the apertures from  \cite{2026AA...706L...4K}, also see Section \ref{sec:met:obsfit}.}
    \label{fig:fit_ch3cn}
\end{figure*}

The fits are shown in Fig. \ref{fig:fit_ch3cn}, derived column densities and approximate abundances relative to solid H$_2$O are listed in Table \ref{tab:coldens_metcyn}. The $\Delta_i$ values were calculated for the spectral sets (see Section \ref{sec:disc:met_fit}). Based on $\Delta_i$ values, there is strong evidence towards the Apolar set for B1-b and Per-emb-27 (g). The fit of Per-emb-27 (w) spectrum shows no CH$_3$CN for the Apolar and Methanol sets. For Per-emb-27 (g) there is considerably less evidence towards HNCO in the methanol environment than for HNCO in the apolar ice layer ($\Delta_i=7.3$). We report $\Delta_i=113.6$--1540.1 for the Null set (only the CH$_3$CN and N$_2$O spectra), therefore the null hypothesis is rejected. More statistical data can be found in Appendix \ref{app:A:fitstat}.

Both HNCO and N$_2$O features are naturally centered within the two peaks of the observational spectra, while CH$_3$CN:H$_2$O is centered on the overlap between 4.42 $\mu$m features and the red wing of $^{13}$CO$_2$ feature. Inclusion of CH$_3$CN decreases the best-fit HNCO column density by 10\% for B1-b and by 18\% for Per-emb-27 (g), and increases the best-fit N$_2$O column density by 11\% and 8\%, respectively. On the other hand, the intensity of the fitted CH$_3$CN feature is close to $1\sigma$ level (see Fig.~\ref{fig:fit_ch3cn}) and changing the aperture for Per-emb-27 removes CH$_3$CN from the fit. That result, together with the 50\% fit uncertainty on CH$_3$CN, indicates that the CH$_3$CN fits are unstable. Also, there are independent CH$_3$CN upper limits derived from the analysis of the 7.27 $\mu$m band towards IRAS 2A (=Per-emb-27) \citep{2024A&A...683A.124R} and B1-c \citep{2024A&A...690A.205C}. Towards IRAS 2A there is an agreement between the upper limits from \cite{2024A&A...683A.124R} and estimates from both this work and \cite{2026A&A...711A.240N}. The CH$_3$CN column density estimates from \cite{2026A&A...711A.240N} towards B1-c exceed the upper limit from \cite{2024A&A...690A.205C} by a factor of 4.5. Considering the overall inconsistency of CH$_3$CN fits we treat the CH$_3$CN column density values listed in Table~\ref{tab:coldens_metcyn} as stringent upper limits.

However, we note that CH$_3$CN improves the best-fit Apolar set AICc against the no-CH$_3$CN fits by $\Delta_i=9.2$ for B1-b and $\Delta_i=9.1$ for Per-emb-27 (g), respectively. This undoubtedly shows that the 2275--2265 cm$^{-1}$ region is underfitted for both sources, but we argue against that CH$_3$CN is the sole carrier of this band.  The influence of the 2275--2265 cm$^{-1}$ feature may diminish the HNCO column densities and increase the N$_2$O column densities for our dataset, but we cannot reliably estimate the column density changes even for cases with little to none H$_2$ emission in $\sim$2270--2265 cm$^{-1}$ range.

\begin{deluxetable*}{cccccccccc}
\tablecaption{Column densities and model comparison for fits with HNCO, N$_2$O and CH$_3$CN.\label{tab:coldens_metcyn}}
\tablewidth{0pt}
\tablehead{
\colhead{Sources} & 
\colhead{\textit{N}(HNCO)} &
\colhead{ $\frac{N(\text{HNCO})}{N(\text{H}_2\text{O})}$} &
\colhead{\textit{N}(N$_2$O)} &
\colhead{ $\frac{N(\text{N}_2\text{O})}{N(\text{H}_2\text{O})}$ }&
\colhead{\textit{N}(CH$_3$CN)} &
\colhead{ $\frac{N(\text{CH$_3$CN})}{N(\text{H}_2\text{O})}$} & 
\multicolumn{3}{c}{$\Delta_i={\rm AICc}_i-{\rm AICc}_{min}$} \\
\colhead{} & 
\colhead{$10^{16}$ cm$^{-2}$} & \colhead{\%} & 
\colhead{$10^{16}$ cm$^{-2}$} & \colhead{\%} & 
\colhead{$10^{16}$ cm$^{-2}$} & \colhead{\%} & 
\colhead{Apolar}& \colhead{Methanol} & \colhead{Water}} 
\startdata
B1-b           & 0.89 $\pm$ 0.07 & 0.08 & 1.04 $\pm$ 0.09 & 0.09 &7.40 $\pm$ 3.93   & 0.66 & 0.0 & 13.7 & 130.5 \\ 
Per-emb-27 (w) & 1.48 $\pm$ 0.10 & 0.08 & 1.27 $\pm$ 0.05 & 0.07 & 0.00 $\pm$ 3.94   & 0 & 0.0 & 59.2 & 329.7 \\ 
Per-emb-27 (g) & 1.91 $\pm$ 0.25 & 0.09 & 3.87 $\pm$ 0.40 & 0.19 & 26.4 $\pm$ 14.0 & 1.28 & 0.0 & 7.3 & 62.9   \\
\enddata
\tablecomments{Column densities are given for the apolar ice model. $\Delta_i=0$ corresponds to the best HNCO ice fitting set (see Section~\ref{sec:met:obsfit}, the best absolute AICc values are listed in Table \ref{app:a:specific_data}). The apertures and designations are taken from \protect\cite{2026AA...706L...4K}. Approximate abundances relative to solid H$_2$O are derived from abundances relative to $^{12}$CO$_2$ using a scale coefficient of 0.28 for CO$_2$/H$_2$O in low mass young stellar objects \protect\citep{2015ARAA..53..541B}.}
\end{deluxetable*}

\subsection{Possible HNCO formation pathways}\label{sec:disc:orig_disc}
In this work we fit 4.42 $\mu$m band previously tentatively assigned to solid interstellar HNCO, performed numerical modelling of HNCO in protostellar objects and analyzed its column density correlations with other species: CO, $^{12}$CO$_2$, OCN$^-$ and N$_2$O. 
HNCO is thought to be a key intermediate to OCN$^-$ with laboratory studies highlighting the ease and effectiveness of HNCO$\to$OCN$^-$ conversion in the H$_2$O-dominated environment \citep{2004CP....302..259R} or in presence of base, i.e. NH$_3$ \citep{2004A&A...415..425V}. Observational evidence suggests that OCN$^-$ forms in CO associated with polar layer of the ice \citep{2004A&A...415..425V,2011ApJ...740..109O,2022ApJ...941...32B}. This hints at a chemical link between CO and HNCO, the latter is supposed to be a key precursor to OCN$^-$ \citep{2004A&A...415..425V,2011A&A...530A..96T}. In addition, \cite{2001ApJ...550.1140H} showed that OCN$^-$ formation requires CO and nitrogen-bearing species, reporting formation pathways through NH$_3$ or~N$_2$. \cite{2018MNRAS.475.1819F} showed that processing of N$_2$-bearing mixtures provides HNCO/OCN$^-$$\ge$1 and a detectable yield of N$_2$O, whereas processing of NH$_3$-bearing mixture does not produce neither HNCO nor N$_2$O. Also, HNCO yield in `H$_2$O-rich' ices is at least by an order of magnitude lower than in `CO-rich ices', as seen from their Table 5. The latter may reflect more efficient survival of HNCO in CO-rich ice: H$^{15}$NCO was observed in H$_2$:CO:$^{15}$N$_2$ ices irradiated at 4~K, while $^{15}$NH$_3$ was not detected and OC$^{15}$N$^-$ could not be unambiguously identified \citep{MD20}.

We report HNCO/OCN$^-$$<$1 (see Fig.~\ref{sec:res:corr:corr}) and suggest that observed HNCO is originated from the apolar CO-dominated layer of the ice. First, the fitting results are consistent with both solid HNCO and solid N$_2$O residing in the apolar layer of the interstellar ice (see Section \ref{sec:res:fit}). Second, there are positive statistically significant correlations between HNCO and both OCN$^-$ and N$_2$O (see panels \textit{c-h} in Fig \ref{fig:corr}). The correlation between HNCO and N$_2$O implies the viability of N$_2$-related HNCO formation pathway and presence of solid N$_2$ in interstellar ice, although the investigation of the exact route is out of the scope of this work. There are two factors that may contribute to observed HNCO/OCN$^-$$<$1: some of HNCO gets deprotonated on formation, contributing to the abundance of OCN$^-$; there may be other precursors to OCN$^-$. Both factors explain the positive correlation between HNCO and OCN$^-$.

Our modeling confirms that, in a gravitational collapse model representing the early stage of protostellar evolution, $\rm HNCO$ can be chemically linked to apolar ice components such as $\rm CO$. The model results show that HNCO is mostly accumulated in the outer CO-rich layers of the ice. As the $\rm HNCO$ formation chain begins with the $\rm CO + OH$ reaction, ice layers rich in $\rm CO$ also contain enhanced abundances of $\rm HNCO$, as seen in the layer-by-layer ice structure (Fig.~\ref{fig:model:lbl}). The barrierless reaction $\rm HNCO + NH_3$, which forms the non-desorbing inert salt $\rm NH_4^+OCN^-$ and is less effective than the hydrogenation reaction $\rm HNCO + H \rightarrow NH_2CO$, however, accounts for a few percent of the total $\rm HNCO$ destruction rate. While $\rm NH_2CO$ can subsequently participate in the formation of formamide ($\rm NH_2CHO$), $\rm NH_4^+OCN^-$ accumulates in ice. We also explored a pathway for solid HNCO formation discussed in \cite{Gerakines_ea2004}. They reported HNCO formation in $\rm HCN : H_2O$ ices at 18~K under energetic processing. Since the authors did not provide kinetic data that can be readily included to the MONACO model, we simplistically included the barrierless reaction HCN~+~H$_2$O~$\rightarrow$~HNCO~+~H$_2$ to the model. This resulted in just 5\% increase in the solid HNCO abundance, owing to the low abundance of solid HCN. 

In summary, the combination of spectral fitting, correlation analysis and modeling reveals that HNCO is not simply an intermediate in OCN$^-$ formation. It likely forms in both polar and apolar environments. While polar HNCO is rapidly converted to OCN$^-$, its apolar counterpart is preserved and co-accumulates with N$_2$O. HNCO in the apolar layer can be further processed to amino acids of astrobiological importance \citep{2015MNRAS.446..439F}.

\section{Conclusions} \label{sec:conclusions}
We presented a systematic analysis of solid HNCO-containing ice mixtures and spectroscopic analysis of HNCO features in open JWST observational data, confirming its presence in interstellar ice. The results of this work can be summarized as follows:
\begin{enumerate}
    \item We obtained the transmission laboratory reference spectra of solid HNCO in various environments providing the dataset for HNCO identification and subsequent study in interstellar ices. We also derived the band strengths of the main HNCO feature ($\nu_2$, 2262 cm$^{-1}$). In addition, we share the dataset of solid HNCO infrared spectra in various molecular environments for temperatures $\geq 10$~K in the 4000--840 cm$^{-1}$ range on Zenodo (DOI: \href{https://zenodo.org/records/22923649?token=eyJhbGciOiJIUzUxMiJ9.eyJpZCI6IjdhYjRjYTBjLWY0OGMtNDk5NC1iNGNjLWY1OTk4NjBlNjc1MSIsImRhdGEiOnt9LCJyYW5kb20iOiI5ODkxNWE3YjFhZTZlZTVhNjQzOGQzZDU5NGE1OTk0MCJ9.zW93j089UXQcRNkOiGZZ1f17cdkcqV2TlG4Dgc-qtwEH-q_E8I3w3b0YnrNc6TDWMtpp9KY0Kp1Gu6Y3pGM87w}{10.5281/zenodo.22923648}).
    \item We used the obtained laboratory data to fit the observational data towards 11 sources and confirmed the presence of solid HNCO in interstellar ice. The results indicate that HNCO predominantly resides in the apolar layer, confirming its origin in CO-rich ices, as proposed by \cite{2018MNRAS.475.1819F}. However, minor inclusions of HNCO in polar or methanol-rich environments cannot be completely excluded. We report HNCO column density in range of (0.43--2.34)\texttimes10$^{16}$ cm$^{-2}$. The HNCO abundance relative to H$_2$O ice is estimated to be approximately in 0.04--0.15\% range.  
    \item Positive correlations between HNCO and N$_2$O show that a significant fraction of observed HNCO is formed in the apolar ice rich in CO and N$_2$, confirming the formation mechanism from \cite{2018MNRAS.475.1819F}. HNCO that originates from the polar layer of the ice is deprotonated to OCN$^-$, supplying the overabundance of OCN$^-$ over HNCO.
    \item Our modeling of the HNCO abundance during gravitational collapse shows that HNCO accumulates in CO-rich ice layers with an abundance of 0.06--0.11\% relative to solid H$_2$O.
    \item We fit 4.41 $\mu$m feature of CH$_3$CN in a spectrum of B1-b and in one of the two spectra of Per-emb-27 (NGC 133 IRAS 2A). The fits yielded $N(\text{CH}_3\text{CN})<7.4\times10^{16}$ cm$^{-2}$ toward B1-b and $N(\text{CH}_3\text{CN})<26.4\times10^{16}$ cm$^{-2}$ toward Per-emb-27. However, since fitted features are at the $1\sigma$ noise level, we treat these values as upper limits, in contrast to \citet{2024A&A...686A..71N,2026A&A...711A.240N} who claimed similar estimates as tentative detection of solid CH$_3$CN. Additionally, our modeling predicts the $\rm CH_3CN$ abundance of the order of $10^{-6}$ relative to water ice.
\end{enumerate}

\section*{Data availability}\label{sec:data_av}

All observational spectra obtained via processing of original MAST data cubes are available on Zenodo (DOI: \href{https://zenodo.org/records/19681305?token=eyJhbGciOiJIUzUxMiIsImlhdCI6MTc3OTEyOTk3NSwiZXhwIjoxNzk4NzYxNTk5fQ.eyJpZCI6IjFjNTg1NTExLTM1YzgtNDZiOC1iNDg5LTFkZGUzMzI0MThhNyIsImRhdGEiOnt9LCJyYW5kb20iOiI4ZDZkMzBmYTdkMmFhZTY0YmIxYjRjN2M2OTJjNTJjNCJ9.smIUL-jYLBeNcKw4jPWDCW4JfH9vjb3KIGLOXNrpC0qeLI42HkPoOpjjWwcJSe8jiBAUK1fj-GrIEG3HkDkYlA}{10.5281/zenodo.19681305}). These spectra cover full NIRSpec range (2.87--5.1 $\mu$m) and can be used by the community to study other ice features. We also share the spectra of HNCO in various environments (DOI: \href{https://zenodo.org/records/22923649?token=eyJhbGciOiJIUzUxMiJ9.eyJpZCI6IjdhYjRjYTBjLWY0OGMtNDk5NC1iNGNjLWY1OTk4NjBlNjc1MSIsImRhdGEiOnt9LCJyYW5kb20iOiI5ODkxNWE3YjFhZTZlZTVhNjQzOGQzZDU5NGE1OTk0MCJ9.zW93j089UXQcRNkOiGZZ1f17cdkcqV2TlG4Dgc-qtwEH-q_E8I3w3b0YnrNc6TDWMtpp9KY0Kp1Gu6Y3pGM87w}{10.5281/zenodo.22923648}).

\section*{}
This work is based on observations made with the NASA/ESA/CSA James Webb Space Telescope. The data were obtained from the Mikulski Archive for Space Telescopes at the Space Telescope Science Institute, which is operated by the Association of Universities for Research in Astronomy, Inc., under NASA contract NAS 5-03127 for JWST. We thank the anonymous reviewers for their insightful comments that helped us to improve the manuscript. This research work is funded by the Russian Science Foundation via 23-12-00315-$\Pi$ agreement.

\begin{figure}[bp]
  \centering
   \includegraphics[width=\hsize]{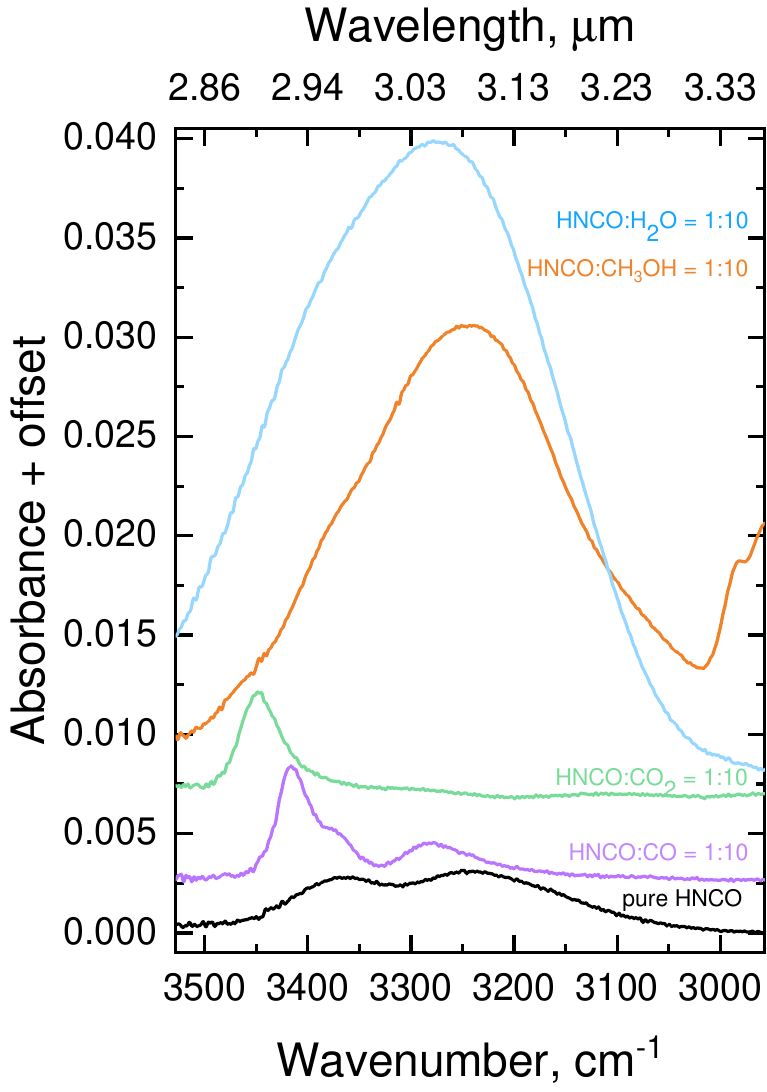}
\caption{Position and profile the NH stretch mode ($\nu_1$) of the HNCO in various mixtures at 10 K: black shows pure HNCO, purple the HNCO:CO, green the HNCO:CO$_2$, cyan the HNCO:H$_2$O, orange the HNCO:CH$_3$OH. The column density of HNCO in all the mixtures equals to 1.1\texttimes10$^{16}$~cm$^{-2}$.}
\label{fig:app:b_specs}
\end{figure}

\begin{figure*}[htbp]
    \centering
    \includegraphics[width=\textwidth]{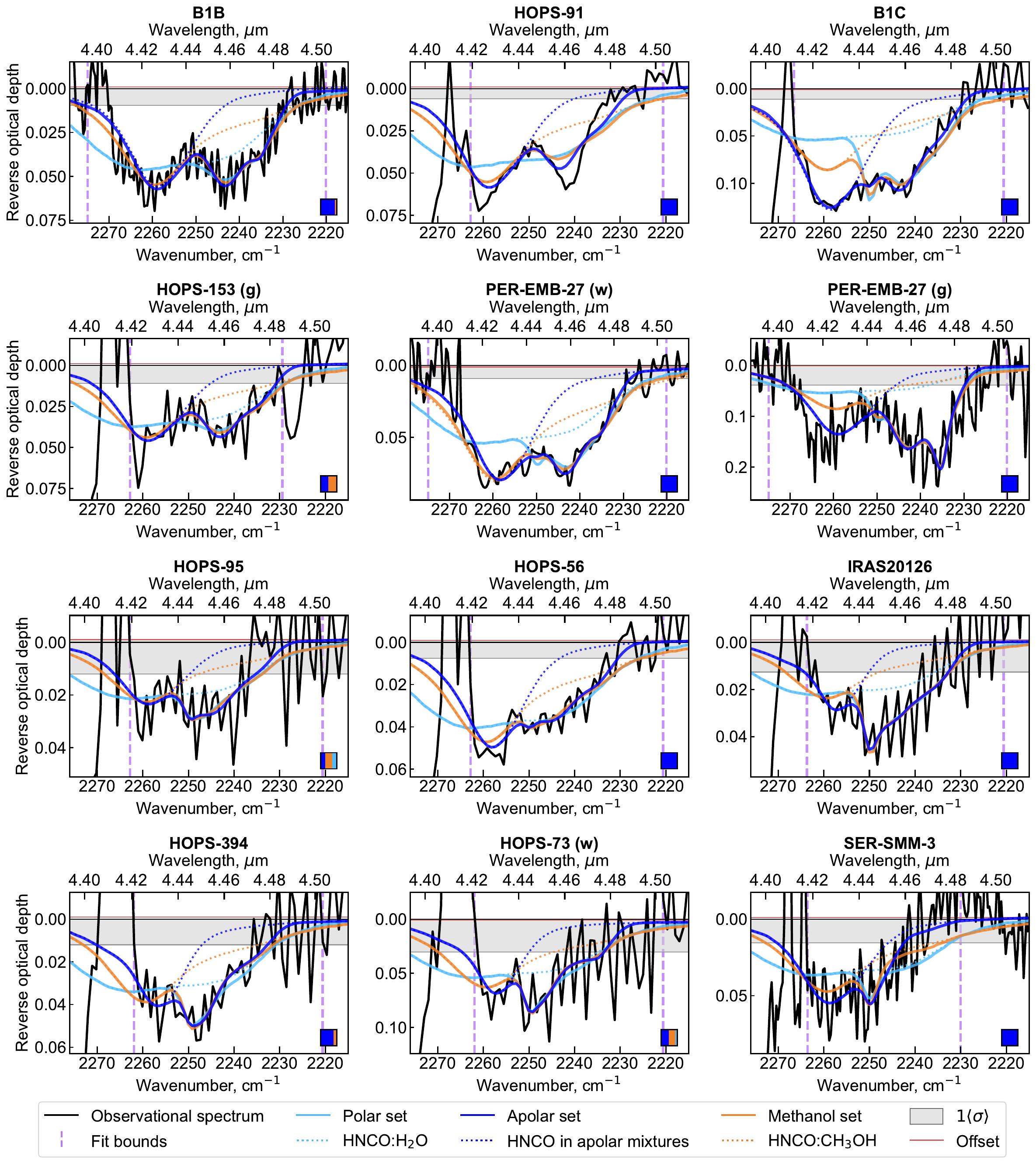}
    \caption{Observation data in 2270--2215 cm$^{-1}$ (4.4--4.55 $\mu$m) region and the best-fit set of spectra. Cyan lines are for HNCO in the Water set, blue --- are HNCO in the Apolar set, orange --- in the Methanol set. Solid lines are the fitted curves, dotted lines are the  The dashed vertical lines indicate the bounds of the fitted range. The letters \textit{g} and \textit{w} denote the colors of the apertures from  \cite{2026AA...706L...4K}, also see Section \ref{sec:met:obsfit}. The squares represent the Akaike weights of the tested spectral sets, more colored space means the higher probability for corresponding set to be the best-approximation (see Section \ref{sec:disc:met_fit}). The blue color represents the Apolar set, green the Methanol set, orange the Water set.}
    \label{fig:app_fit}
\end{figure*}

\appendix 
In Appendix \ref{app:A:fitstat} we comment the fit statistics. Appendix \ref{app:B:tablesandplots} includes additional tables and figures: shared parameters of the fits (Table \ref{app:a:shared_data}); fit results for fits with HNCO, N$_2$O and CH$_3$CN (Table \ref{app:a:specific_data}); position and profile the NH stretch mode ($\nu_1$)
of the HNCO in various mixtures at 10 K (Figure \ref{fig:app:b_specs}); the fitted sets of spectra and observational data in 2270--2215 cm$^{-1}$ (4.4--4.55 $\mu$m) region (Figure \ref{fig:app_fit}).

\section{Fit statistics}\label{app:A:fitstat}
In this section we motivate the use of AICc and provide the supporting statistical data for the fits in Tables~\ref{tab:coldens} and \ref{tab:coldens_metcyn}. \cite{jabi2} argue that AIC is appropriate when models are treated as approximations to a complex reality with tapering effects, whereas BIC (Bayesian information criterion) is favored when a few large effects dominate and a (quasi-)true model may be in the candidate set. Indeed, interstellar ice represent a complex reality as they may display gradients of temperature or species composition towards the line of sight. In this paper we utilized a rather simplistic ice composition only to determine the most likely environment of HNCO, therefore the AIC was chosen over BIC. In turn, the corrected AIC (AICc) was chosen due to the $n/k<40$ \citep{jabi2}, where $n$ is the sample size and $k$ is the number of free parameters. There are 7, 8, 7 and 5 free parameters for Polar, Apolar, Methanol and Null sets for no-CH$_3$CN fits, respectively. An addition of CH$_3$CN increases the amount of free parameters by one. The number of points $N_{pts}$ in the fitted wavenumber ranges and number of fitted components $k$ for our fits are listed in Table \ref{app:a:shared_data} and \ref{app:a:specific_data}, respectively. The goodness-of-fit was assessed with the Anderson–Darling statistic $A^2$ of the standardized residuals against $\mathcal{N}(0,1)$ \citep{Stephens1974EDFSF}. Following \cite{2012msma.book.....F}, the significance of $A^2$ is obtained by bootstrap resampling rather than from tabulated EDF probabilities, which do not apply when the model is estimated from the same sample. Values of $A^2$ and the bootstrap $p_{\mathrm{AD}}$ for the preferred fits are listed in Table~\ref{app:a:specific_data}. All preferred models have $p_{\mathrm{AD}}>0.05$. The $A^2$ values themselves are $\lesssim 2.49$, which is Stephens’ 5\% point for a fully specified $\mathcal{N}(0,1)$, except for HOPS-91 and Ser-SMM-3, and $\chi^2_{\mathrm{red}}$ is in range of 1.10--1.48 (2.32 for HOPS-91). The residuals are therefore compatible with the adopted noise model at the level required for the weighted fit. 

\begin{deluxetable}{ccccc}[tbp]
\section{Tables and plots}\label{app:B:tablesandplots}
\tablecaption{Shared parameters of the fits}\label{app:a:shared_data}
\tablewidth{0pt}
\tablehead{
\colhead{Source} &
\colhead{$N_{pts}$} &
\colhead{Range} &
\colhead{$\langle\sigma\rangle$} &
\colhead{$\tilde{\sigma}$} \\
\colhead{} &
\colhead{} &
\colhead{cm$^{-1}$} &
\colhead{10$^{-3}$} &
\colhead{10$^{-3}$}
}
\startdata
B1-b           & 164 & 2274.9--2220.1 & 9.5  & 9.0  \\
HOPS-91        & 48  & 2262.8--2220.6 & 6.0  & 6.0  \\
B1-c           & 52  & 2266.5--2220.6 & 11.3 & 9.7  \\
HOPS-153 (g)   & 38  & 2262.8--2229.4 & 10.9 & 10.1 \\
Per-emb-27 (w) & 164 & 2274.9--2220.1 & 9.0  & 6.5  \\
Per-emb-27 (g) & 164 & 2274.9--2220.1 & 38.8 & 31.0 \\
HOPS-95        & 48  & 2262.8--2220.6 & 12.1 & 10.5 \\
HOPS-56        & 48  & 2262.8--2220.6 & 7.6  & 6.4  \\
IRAS20126      & 49  & 2263.7--2220.6 & 12.6 & 12.7 \\
HOPS-394       & 47  & 2261.9--2220.6 & 12.0 & 10.2 \\
HOPS-73 (w)    & 47  & 2261.9--2220.6 & 30.4 & 28.7 \\
Ser-SMM-3      & 101 & 2263.6--2230.0 & 15.3 & 15.5 \\
\enddata
\tablecomments{$\langle\sigma\rangle$ and $\tilde\sigma$ correspond to mean and median standard deviation of the data in fitted observational spectra (see Section \ref{sec:disc:met_fit}).}
\end{deluxetable}
\begin{deluxetable}{ccccccc}[htbp]
\tablecaption{Fit results for fits with HNCO, N$_2$O and CH$_3$CN.}\label{app:a:specific_data}
\tablewidth{0pt}
\tablehead{
\colhead{Source} &
\colhead{Set} &
\colhead{$\chi^2$} &
\colhead{$\chi^2_{\rm red}$} &
\colhead{$A^2$} &
\colhead{$p_{AD}$} &
\colhead{AICc}}
\startdata
\hline
\multicolumn{7}{c}{Fits with HNCO and N$_2$O} \\
\hline
B1-b           & A & 205.9 & 1.32 & 2.16 & 0.48 & 222.9 \\
HOPS-91        & A & 92.88 & 2.32 & 2.78 & 0.32 & 112.6 \\
B1-c           & A & 49.62 & 1.13 & 0.97 & 0.30 & 69.0 \\
HOPS-153 (g)   & M & 41.48 & 1.34 & 0.25 & 0.95 & 59.2 \\
               & A & 38.28 & 1.28 & 0.32 & 0.93 & 59.3 \\
Per-emb-27 (w) & A & 172.0 & 1.10 & 0.54 & 0.83 & 189.0 \\
Per-emb-27 (g) & A & 229.2 & 1.47 & 1.78 & 0.51 & 246.1 \\
HOPS-95        & M & 56.15 & 1.37 & 0.90 & 0.45 & 73.0 \\
               & A & 54.17 & 1.35 & 0.66 & 0.68 & 73.9 \\
               & W & 57.23 & 1.40 & 0.81 & 0.55 & 74.0 \\
HOPS-56        & A & 46.11 & 1.15 & 0.38 & 0.85 & 65.8 \\
IRAS20126      & A & 51.75 & 1.26 & 0.97 & 0.32 & 71.4 \\
HOPS-394       & A & 46.86 & 1.20 & 0.47 & 0.73 & 66.7 \\
HOPS-73 (w)    & A & 50.56 & 1.30 & 0.41 & 0.81 & 70.4 \\
               & M & 53.93 & 1.35 & 0.26 & 0.96 & 70.8 \\
Ser-SMM-3      & A & 137.6 & 1.48 & 2.59 & 0.15 & 155.2 \\
\hline
\multicolumn{7}{c}{Fits with HNCO, N$_2$O and CH$_3$CN} \\
\hline
B1-b           & A & 194.6 & 1.26 & 1.29 & 0.47 & 213.7 \\
Per-emb-27 (w) & A & 172.0 & 1.11 & 0.54 & 0.79 & 191.2 \\
Per-emb-27 (g) & A & 217.8 & 1.41 & 1.11 & 0.56 & 237.0 \\
\enddata
\tablecomments{The preferred set is that with the lowest AICc (see Section \ref{sec:disc:met_fit}); a second-best set is listed when $\Delta_i=\text{AICc}_i-\text{AICc}_{min} \le 2$. See Table \ref{tab:coldens} for $\Delta_i$ values. Anderson--Darling bootstrap $(A^2,p_{AD})$ tests of residual normality is considered passed at $\alpha = 0.05$ ($p > 0.05$).}
\end{deluxetable}

\clearpage

\bibliography{Nakibov}{}

@ARTICLE{1972ApJ...177..619S,
       author = {{Snyder}, Lewis E. and {Buhl}, David},
        title = "{Interstellar Isocyanic Acid}",
      journal = {\apj},
         year = 1972,
        month = nov,
       volume = {177},
        pages = {619},
          doi = {10.1086/151739},
       adsurl = {https://ui.adsabs.harvard.edu/abs/1972ApJ...177..619S}
}

@ARTICLE{2015MNRAS.446..439F,
       author = {{Fedoseev}, G. and {Ioppolo}, S. and {Zhao}, D. and {Lamberts}, T. and {Linnartz}, H.},
        title = "{Low-temperature surface formation of NH$_{3}$ and HNCO: hydrogenation of nitrogen atoms in CO-rich interstellar ice analogues}",
      journal = {\mnras},
         year = 2015,
        month = jan,
       volume = {446},
       number = {1},
        pages = {439-448},
          doi = {10.1093/mnras/stu2028},
archivePrefix = {arXiv},
       eprint = {1705.09184},
 primaryClass = {astro-ph.GA},
       adsurl = {https://ui.adsabs.harvard.edu/abs/2015MNRAS.446..439F}
}

@ARTICLE{2004A&A...415..425V,
       author = {{van Broekhuizen}, F.~A. and {Keane}, J.~V. and {Schutte}, W.~A.},
        title = "{A quantitative analysis of OCN$^{-}$ formation in interstellar ice analogs}",
      journal = {\aap},
         year = 2004,
        month = feb,
       volume = {415},
        pages = {425-436},
          doi = {10.1051/0004-6361:20034161},
archivePrefix = {arXiv},
       eprint = {astro-ph/0311617},
 primaryClass = {astro-ph},
       adsurl = {https://ui.adsabs.harvard.edu/abs/2004A&A...415..425V}
}

@ARTICLE{2024ApJ...972L..10O,
       author = {{Ozhiganov}, Maksim and {Medvedev}, Mikhail and {Karteyeva}, Varvara and {Nakibov}, Ruslan and {Sapunova}, Uliana and {Krushinsky}, Vadim and {Stepanova}, Ksenia and {Tryastsina}, Anastasia and {Gorkovenko}, Aleksandr and {Fedoseev}, Gleb and {Vasyunin}, Anton},
        title = "{Infrared Spectra of Solid HCN Embedded in Various Molecular Environments for Comparison with the Data Obtained with JWST}",
      journal = {\apjl},
         year = 2024,
        month = sep,
       volume = {972},
       number = {1},
          eid = {L10},
        pages = {L10},
          doi = {10.3847/2041-8213/ad6d5c},
archivePrefix = {arXiv},
       eprint = {2408.08166},
 primaryClass = {astro-ph.GA},
       adsurl = {https://ui.adsabs.harvard.edu/abs/2024ApJ...972L..10O}
}

@ARTICLE{2004CP....302..259R,
       author = {{Raunier}, S{\'e}bastien and {Chiavassa}, Thierry and {Marinelli}, Francis and {Aycard}, Jean-Pierre},
        title = "{Experimental and theoretical study on the spontaneous formation of OCN $^{-}$ ion: reactivity between HNCO and NH $_{3}$/H $_{2}$O environment at low temperature}",
      journal = {CP},
         year = 2004,
        month = jul,
       volume = {302},
       number = {1},
        pages = {259-264},
          doi = {10.1016/j.chemphys.2004.04.013},
       adsurl = {https://ui.adsabs.harvard.edu/abs/2004CP....302..259R}
}

@ARTICLE{2018MNRAS.475.1819F,
       author = {{Fedoseev}, G. and {Scir{\`e}}, C. and {Baratta}, G.~A. and {Palumbo}, M.~E.},
        title = "{Cosmic ray processing of N$_{2}$-containing interstellar ice analogues at dark cloud conditions}",
      journal = {\mnras},
         year = 2018,
        month = apr,
       volume = {475},
       number = {2},
        pages = {1819-1828},
          doi = {10.1093/mnras/stx3302},
       adsurl = {https://ui.adsabs.harvard.edu/abs/2018MNRAS.475.1819F}
}

@ARTICLE{2018MNRAS.474.2796Q,
       author = {{Qu{\'e}nard}, David and {Jim{\'e}nez-Serra}, Izaskun and {Viti}, Serena and {Holdship}, Jonathan and {Coutens}, Audrey},
        title = "{Chemical modelling of complex organic molecules with peptide-like bonds in star-forming regions}",
      journal = {\mnras},
         year = 2018,
        month = feb,
       volume = {474},
       number = {2},
        pages = {2796-2812},
          doi = {10.1093/mnras/stx2960},
archivePrefix = {arXiv},
       eprint = {1711.05184},
 primaryClass = {astro-ph.SR},
       adsurl = {https://ui.adsabs.harvard.edu/abs/2018MNRAS.474.2796Q}
}

@ARTICLE{2025MNRAS.537.2918G,
       author = {{Gerakines}, Perry A. and {Materese}, Christopher K. and {Hudson}, Reggie L.},
        title = "{Establishing an accurate infrared band strength for the cyanate ion in interstellar ices}",
      journal = {\mnras},
         year = 2025,
        month = mar,
       volume = {537},
       number = {3},
        pages = {2918-2924},
          doi = {10.1093/mnras/staf192},
       adsurl = {https://ui.adsabs.harvard.edu/abs/2025MNRAS.537.2918G}
}

@ARTICLE{2010ApJ...725.2101Q,
       author = {{Quan}, Donghui and {Herbst}, Eric and {Osamura}, Yoshihiro and {Roueff}, Evelyne},
        title = "{Gas-grain Modeling of Isocyanic Acid (HNCO), Cyanic Acid (HOCN), Fulminic Acid (HCNO), and Isofulminic Acid (HONC) in Assorted Interstellar Environments}",
      journal = {\apj},
         year = 2010,
        month = dec,
       volume = {725},
       number = {2},
        pages = {2101-2109},
          doi = {10.1088/0004-637X/725/2/2101},
       adsurl = {https://ui.adsabs.harvard.edu/abs/2010ApJ...725.2101Q}
}

@ARTICLE{2024A&A...690A.271B,
       author = {{Biver}, N. and {Bockel{\'e}e-Morvan}, D. and {Handzlik}, B. and {Sandqvist}, Aa. and {Boissier}, J. and {Drozdovskaya}, M.~N. and {Moreno}, R. and {Crovisier}, J. and {Lis}, D.~C. and {Cordiner}, M. and {Milam}, S. and {Roth}, N.~X. and {Bonev}, B.~P. and {Dello Russo}, N. and {Vervack}, R. and {Opitom}, C. and {Kawakita}, H.},
        title = "{Chemical composition of comets C/2021 A1 (Leonard) and C/2022 E3 (ZTF) from radio spectroscopy and the abundance of HCOOH and HNCO in comets}",
      journal = {\aap},
         year = 2024,
        month = oct,
       volume = {690},
          eid = {A271},
        pages = {A271},
          doi = {10.1051/0004-6361/202450921},
archivePrefix = {arXiv},
       eprint = {2408.10759},
 primaryClass = {astro-ph.EP},
       adsurl = {https://ui.adsabs.harvard.edu/abs/2024A&A...690A.271B}
}

@ARTICLE{2022A&A...665A.131L,
       author = {{Lin}, Y. and {Spezzano}, S. and {Sipil{\"a}}, O. and {Vasyunin}, A. and {Caselli}, P.},
        title = "{Multiline observations of CH$_{3}$OH, c-C$_{3}$H$_{2}$, and HNCO toward L1544. Dissecting the core structure with chemical differentiation}",
      journal = {\aap},
         year = 2022,
        month = sep,
       volume = {665},
          eid = {A131},
        pages = {A131},
          doi = {10.1051/0004-6361/202243657},
archivePrefix = {arXiv},
       eprint = {2205.09806},
 primaryClass = {astro-ph.GA},
       adsurl = {https://ui.adsabs.harvard.edu/abs/2022A&A...665A.131L}
}

@ARTICLE{2016MNRAS.460.4297F,
       author = {{Fedoseev}, G. and {Chuang}, K.-J. and {van Dishoeck}, E.~F. and {Ioppolo}, S. and {Linnartz}, H.},
        title = "{Simultaneous hydrogenation and UV-photolysis experiments of NO in CO-rich interstellar ice analogues; linking HNCO, OCN$^{-}$, NH$_{2}$CHO, and NH$_{2}$OH}",
      journal = {\mnras},
         year = 2016,
        month = aug,
       volume = {460},
       number = {4},
        pages = {4297-4309},
          doi = {10.1093/mnras/stw1028},
archivePrefix = {arXiv},
       eprint = {1706.00736},
 primaryClass = {astro-ph.GA},
       adsurl = {https://ui.adsabs.harvard.edu/abs/2016MNRAS.460.4297F}
}

@ARTICLE{2017MNRAS.467.2723P,
       author = {{Palau}, Aina and {Walsh}, Catherine and {S{\'a}nchez-Monge}, {\'A}lvaro and {Girart}, Josep M. and {Cesaroni}, Riccardo and {Jim{\'e}nez-Serra}, Izaskun and {Fuente}, Asunci{\'o}n and {Zapata}, Luis A. and {Neri}, Roberto},
        title = "{Complex organic molecules tracing shocks along the outflow cavity in the high-mass protostar IRAS 20126+4104}",
      journal = {\mnras},
         year = 2017,
        month = may,
       volume = {467},
       number = {3},
        pages = {2723-2752},
          doi = {10.1093/mnras/stx004},
archivePrefix = {arXiv},
       eprint = {1701.04802},
 primaryClass = {astro-ph.SR},
       adsurl = {https://ui.adsabs.harvard.edu/abs/2017MNRAS.467.2723P}
}

@ARTICLE{2019MNRAS.483.2014H,
       author = {{Hern{\'a}ndez-G{\'o}mez}, Antonio and {Sahnoun}, Emna and {Caux}, Emmanuel and {Wiesenfeld}, Laurent and {Loinard}, Laurent and {Bottinelli}, Sandrine and {Hammami}, Kamel and {Menten}, Karl M.},
        title = "{Modelling the abundance structure of isocyanic acid (HNCO) towards the low-mass solar type protostar IRAS 16293-2422}",
      journal = {\mnras},
         year = 2019,
        month = feb,
       volume = {483},
       number = {2},
        pages = {2014-2030},
          doi = {10.1093/mnras/sty2971},
archivePrefix = {arXiv},
       eprint = {1811.00766},
 primaryClass = {astro-ph.SR},
       adsurl = {https://ui.adsabs.harvard.edu/abs/2019MNRAS.483.2014H}
}

@ARTICLE{2010A&A...516A.105M,
       author = {{Marcelino}, N. and {Br{\"u}nken}, S. and {Cernicharo}, J. and {Quan}, D. and {Roueff}, E. and {Herbst}, E. and {Thaddeus}, P.},
        title = "{The puzzling behavior of HNCO isomers in molecular clouds}",
      journal = {\aap},
         year = 2010,
        month = jun,
       volume = {516},
          eid = {A105},
        pages = {A105},
          doi = {10.1051/0004-6361/200913806},
       adsurl = {https://ui.adsabs.harvard.edu/abs/2010A&A...516A.105M}
}

@ARTICLE{2009A&A...494L..13O,
       author = {{{\"O}berg}, K.~I. and {Bottinelli}, S. and {van Dishoeck}, E.~F.},
        title = "{Cold gas as an ice diagnostic toward low mass protostars}",
      journal = {\aap},
         year = 2009,
        month = feb,
       volume = {494},
       number = {2},
        pages = {L13-L16},
          doi = {10.1051/0004-6361:200811228},
archivePrefix = {arXiv},
       eprint = {0901.1019},
 primaryClass = {astro-ph.SR},
       adsurl = {https://ui.adsabs.harvard.edu/abs/2009A&A...494L..13O}
}

@ARTICLE{2015ARAA..53..541B,
       author = {{Boogert}, A.~C. Adwin and {Gerakines}, Perry A. and {Whittet}, Douglas C.~B.},
        title = "{Observations of the icy universe.}",
      journal = {\araa},
         year = 2015,
        month = aug,
       volume = {53},
        pages = {541-581},
          doi = {10.1146/annurev-astro-082214-122348},
archivePrefix = {arXiv},
       eprint = {1501.05317},
 primaryClass = {astro-ph.GA},
       adsurl = {https://ui.adsabs.harvard.edu/abs/2015ARA&A..53..541B}
}

@ARTICLE{2014ApJ...788...19J,
       author = {{Jim{\'e}nez-Escobar}, A. and {Giuliano}, B.~M. and {Mu{\~n}oz Caro}, G.~M. and {Cernicharo}, J. and {Marcelino}, N.},
        title = "{Investigation of HNCO Isomer Formation in Ice Mantles by UV and Thermal Processing: An Experimental Approach}",
      journal = {\apj},
         year = 2014,
        month = jun,
       volume = {788},
       number = {1},
          eid = {19},
        pages = {19},
          doi = {10.1088/0004-637X/788/1/19},
archivePrefix = {arXiv},
       eprint = {1407.1242},
 primaryClass = {astro-ph.IM},
       adsurl = {https://ui.adsabs.harvard.edu/abs/2014ApJ...788...19J}
}

@ARTICLE{2010A&A...510A..85T,
       author = {{Tideswell}, D.~M. and {Fuller}, G.~A. and {Millar}, T.~J. and {Markwick}, A.~J.},
        title = "{The abundance of HNCO and its use as a diagnostic of environment}",
      journal = {\aap},
         year = 2010,
        month = feb,
       volume = {510},
          eid = {A85},
        pages = {A85},
          doi = {10.1051/0004-6361/200810820},
archivePrefix = {arXiv},
       eprint = {0909.4337},
 primaryClass = {astro-ph.GA},
       adsurl = {https://ui.adsabs.harvard.edu/abs/2010A&A...510A..85T}
}

@ARTICLE{2011ApJ...743...60H,
       author = {{Halfen}, D.~T. and {Ilyushin}, V. and {Ziurys}, L.~M.},
        title = "{Formation of Peptide Bonds in Space: A Comprehensive Study of Formamide and Acetamide in Sgr B2(N)}",
      journal = {\apj},
         year = 2011,
        month = dec,
       volume = {743},
       number = {1},
          eid = {60},
        pages = {60},
          doi = {10.1088/0004-637X/743/1/60},
       adsurl = {https://ui.adsabs.harvard.edu/abs/2011ApJ...743...60H}
}

@ARTICLE{2015MNRAS.451.2145B,
       author = {{Bouilloud}, M. and {Fray}, N. and {B{\'e}nilan}, Y. and {Cottin}, H. and {Gazeau}, M.-C. and {Jolly}, A.},
        title = "{Bibliographic review and new measurements of the infrared band strengths of pure molecules at 25 K: H$_{2}$O, CO$_{2}$, CO, CH$_{4}$, NH$_{3}$, CH$_{3}$OH, HCOOH and H$_{2}$CO}",
      journal = {\mnras},
         year = 2015,
        month = aug,
       volume = {451},
       number = {2},
        pages = {2145-2160},
          doi = {10.1093/mnras/stv1021},
       adsurl = {https://ui.adsabs.harvard.edu/abs/2015MNRAS.451.2145B}
}

@ARTICLE{2023MNRAS.522.3145G,
       author = {{Gerakines}, Perry A. and {Materese}, Christopher K. and {Hudson}, Reggie L.},
        title = "{Carbon monoxide ices - a semicentennial review and update for crystalline CO along with the first IR spectrum and band strength for amorphous CO}",
      journal = {\mnras},
         year = 2023,
        month = jun,
       volume = {522},
       number = {2},
        pages = {3145-3162},
          doi = {10.1093/mnras/stad1164},
       adsurl = {https://ui.adsabs.harvard.edu/abs/2023MNRAS.522.3145G}
}

@ARTICLE{2015ApJ...808L..40G,
       author = {{Gerakines}, Perry A. and {Hudson}, Reggie L.},
        title = "{First Infrared Band Strengths for Amorphous CO$_{2}$, an Overlooked Component of Interstellar Ices}",
      journal = {\apjl},
         year = 2015,
        month = aug,
       volume = {808},
       number = {2},
          eid = {L40},
        pages = {L40},
          doi = {10.1088/2041-8205/808/2/L40},
       adsurl = {https://ui.adsabs.harvard.edu/abs/2015ApJ...808L..40G}
}

@ARTICLE{2024ApJ...970..108H,
       author = {{Hudson}, Reggie L. and {Gerakines}, Perry A. and {Yarnall}, Yukiko Y.},
        title = "{Infrared Spectroscopic and Physical Properties of Methanol Ices{\textemdash}Reconciling the Conflicting Published Band Strengths of an Important Interstellar Solid}",
      journal = {\apj},
         year = 2024,
        month = aug,
       volume = {970},
       number = {2},
          eid = {108},
        pages = {108},
          doi = {10.3847/1538-4357/ad47a5},
       adsurl = {https://ui.adsabs.harvard.edu/abs/2024ApJ...970..108H}
}

@ARTICLE{2023A&A...677A..13S,
       author = {{Slavicinska}, K. and {Rachid}, M.~G. and {Rocha}, W.~R.~M. and {Chuang}, K.-J. and {van Dishoeck}, E.~F. and {Linnartz}, H.},
        title = "{The hunt for formamide in interstellar ices. A toolkit of laboratory infrared spectra in astronomically relevant ice mixtures and comparisons to ISO, Spitzer, and JWST observations}",
      journal = {\aap},
         year = 2023,
        month = sep,
       volume = {677},
          eid = {A13},
        pages = {A13},
          doi = {10.1051/0004-6361/202346996},
archivePrefix = {arXiv},
       eprint = {2307.04790},
 primaryClass = {astro-ph.GA},
       adsurl = {https://ui.adsabs.harvard.edu/abs/2023A&A...677A..13S}
}

@ARTICLE{2026AA...706L...4K,
       author = {{Karteyeva}, V. and {Nakibov}, R. and {Petrashkevich}, I. and {Medvedev}, M. and {Vasyunin}, A.},
        title = "{Identification of solid N$_{2}$O in interstellar ices using open JWST data}",
      journal = {\aap},
         year = 2026,
        month = jan,
       volume = {706},
          eid = {L4},
        pages = {L4},
          doi = {10.1051/0004-6361/202557939},
       adsurl = {https://ui.adsabs.harvard.edu/abs/2026A&A...706L...4K}
}

@ARTICLE{2018MNRAS.480.3628L,
       author = {{Ligterink}, N.~F.~W. and {Terwisscha van Scheltinga}, J. and {Taquet}, V. and {J{\o}rgensen}, J.~K. and {Cazaux}, S. and {van Dishoeck}, E.~F. and {Linnartz}, H.},
        title = "{The formation of peptide-like molecules on interstellar dust grains}",
      journal = {\mnras},
         year = 2018,
        month = nov,
       volume = {480},
       number = {3},
        pages = {3628-3643},
          doi = {10.1093/mnras/sty2066},
archivePrefix = {arXiv},
       eprint = {1808.00742},
 primaryClass = {astro-ph.SR},
       adsurl = {https://ui.adsabs.harvard.edu/abs/2018MNRAS.480.3628L}
}

@ARTICLE{2023ApJ...949...89X,
       author = {{Xie}, Jinjin and {Li}, Juan and {Wang}, Junzhi and {Liu}, Shu and {Yang}, Kai and {Quan}, Donghui and {Zheng}, Siqi and {Li}, Yuqiang and {Wu}, Jingwen and {Duan}, Yan and {Li}, Di},
        title = "{Imaging Molecular Outflow in Massive Star-forming Regions with HNCO Lines}",
      journal = {\apj},
         year = 2023,
        month = jun,
       volume = {949},
       number = {2},
          eid = {89},
        pages = {89},
          doi = {10.3847/1538-4357/acc83f},
archivePrefix = {arXiv},
       eprint = {2303.14866},
 primaryClass = {astro-ph.GA},
       adsurl = {https://ui.adsabs.harvard.edu/abs/2023ApJ...949...89X}
}

@ARTICLE{2013A&A...555A..13M,
       author = {{Mispelaer}, F. and {Theul{\'e}}, P. and {Aouididi}, H. and {Noble}, J. and {Duvernay}, F. and {Danger}, G. and {Roubin}, P. and {Morata}, O. and {Hasegawa}, T. and {Chiavassa}, T.},
        title = "{Diffusion measurements of CO, HNCO, H$_{2}$CO, and NH$_{3}$ in amorphous water ice}",
      journal = {\aap},
         year = 2013,
        month = jul,
       volume = {555},
          eid = {A13},
        pages = {A13},
          doi = {10.1051/0004-6361/201220691},
       adsurl = {https://ui.adsabs.harvard.edu/abs/2013A&A...555A..13M}
}

@ARTICLE{2011A&A...530A..96T,
       author = {{Theule}, P. and {Duvernay}, F. and {Ilmane}, A. and {Hasegawa}, T. and {Morata}, O. and {Coussan}, S. and {Danger}, G. and {Chiavassa}, T.},
        title = "{Kinetics of the OCN$^{-}$ and HOCN formation from the HNCO + H$_{2}$O thermal reaction in interstellar ice analogs}",
      journal = {\aap},
         year = 2011,
        month = jun,
       volume = {530},
          eid = {A96},
        pages = {A96},
          doi = {10.1051/0004-6361/201016051},
       adsurl = {https://ui.adsabs.harvard.edu/abs/2011A&A...530A..96T}
}

@ARTICLE{2002AcSpA..58...73L,
       author = {{Lowenthal}, M.~S. and {Khanna}, R.~K. and {Moore}, Marla H.},
        title = "{Infrared spectrum of solid isocyanic acid (HNCO): vibrational assignments and integrated band intensities}",
      journal = {    AcSpA    },
         year = 2002,
        month = jan,
       volume = {58},
       number = {1},
        pages = {73-78},
          doi = {10.1016/S1386-1425(01)00524-8},
       adsurl = {https://ui.adsabs.harvard.edu/abs/2002AcSpA..58...73L}
}

@article{karteyeva2026infrared,
  title={Infrared spectra of methane-containing ice mixtures for JWST data analyses},
  author={{Karteyeva}, V. and Nakibov, Ruslan and Petrashkevich, Igor and Medvedev, Mikhail and Vasyunin, Anton},
  journal={A\&A},
  doi={10.1051/0004-6361/202558069},
  volume={707},
  pages={A293},
  year={2026},
  publisher={EDP Sciences}
}

@article{Cleveland_1979,
author = {William S. Cleveland},
title = {Robust Locally Weighted Regression and Smoothing Scatterplots},
journal = {JASA},
volume = {74},
number = {368},
pages = {829--836},
year = {1979},
publisher = {ASA Website},
doi = {10.1080/01621459.1979.10481038},
URL = {https://www.tandfonline.com/doi/abs/10.1080/01621459.1979.10481038},
eprint = {https://www.tandfonline.com/doi/pdf/10.1080/01621459.1979.10481038}
}

@ARTICLE{2011ApJ...729...92B,
       author = {{Boogert}, A.~C.~A. and {Huard}, T.~L. and {Cook}, A.~M. and {Chiar}, J.~E. and {Knez}, C. and {Decin}, L. and {Blake}, G.~A. and {Tielens}, A.~G.~G.~M. and {van Dishoeck}, E.~F.},
        title = "{Ice and Dust in the Quiescent Medium of Isolated Dense Cores}",
      journal = {\apj},
         year = 2011,
        month = mar,
       volume = {729},
       number = {2},
          eid = {92},
        pages = {92},
          doi = {10.1088/0004-637X/729/2/92},
archivePrefix = {arXiv},
       eprint = {1101.2486},
 primaryClass = {astro-ph.GA},
       adsurl = {https://ui.adsabs.harvard.edu/abs/2011ApJ...729...92B}
}

@ARTICLE{Brunken,
       author = {{Brunken}, N.~G.~C. and {van Dishoeck}, E.~F. and {Slavicinska}, K. and {le Gouellec}, V.~J.~M. and {Rocha}, W.~R.~M. and {Francis}, L. and {Tychoniec}, L. and {van Gelder}, M.~L. and {Navarro}, M.~G. and {Boogert}, A.~C.~A. and {Kavanagh}, P.~J. and {Nazari}, P. and {Greene}, T. and {Ressler}, M.~E. and {Majumdar}, L.},
        title = "{JOYS+ study of solid-state $^{12}$C/$^{13}$C isotope ratios in protostellar envelopes: Observations of CO and CO$_{2}$ ice with the James Webb Space Telescope}",
      journal = {\aap},
         year = 2024,
        month = dec,
       volume = {692},
          eid = {A163},
        pages = {A163},
          doi = {10.1051/0004-6361/202451794},
archivePrefix = {arXiv},
       eprint = {2409.17237},
 primaryClass = {astro-ph.SR},
       adsurl = {https://ui.adsabs.harvard.edu/abs/2024A&A...692A.163B}
}

@ARTICLE{2022ApJ...941...32B,
       author = {{Boogert}, A.~C.~A. and {Brewer}, K. and {Brittain}, A. and {Emerson}, K.~S.},
        title = "{Survey of Ices toward Massive Young Stellar Objects. I. OCS, CO, OCN$^{-}$, and CH$_{3}$OH}",
      journal = {\apj},
         year = 2022,
        month = dec,
       volume = {941},
       number = {1},
          eid = {32},
        pages = {32},
          doi = {10.3847/1538-4357/ac9b4a},
archivePrefix = {arXiv},
       eprint = {2210.12639},
 primaryClass = {astro-ph.GA},
       adsurl = {https://ui.adsabs.harvard.edu/abs/2022ApJ...941...32B}
}

@ARTICLE{2001ApJ...550.1140H,
       author = {{Hudson}, R.~L. and {Moore}, M.~H. and {Gerakines}, P.~A.},
        title = "{The Formation of Cyanate Ion (OCN$^{-}$) in Interstellar Ice Analogs}",
      journal = {\apj},
         year = 2001,
        month = apr,
       volume = {550},
       number = {2},
        pages = {1140-1150},
          doi = {10.1086/319799},
       adsurl = {https://ui.adsabs.harvard.edu/abs/2001ApJ...550.1140H}
}

@ARTICLE{2011ApJ...740..109O,
       author = {{{\"O}berg}, Karin I. and {Boogert}, A.~C. Adwin and {Pontoppidan}, Klaus M. and {van den Broek}, Saskia and {van Dishoeck}, Ewine F. and {Bottinelli}, Sandrine and {Blake}, Geoffrey A. and {Evans}, II, Neal J.},
        title = "{The Spitzer Ice Legacy: Ice Evolution from Cores to Protostars}",
      journal = {\apj},
         year = 2011,
        month = oct,
       volume = {740},
       number = {2},
          eid = {109},
        pages = {109},
          doi = {10.1088/0004-637X/740/2/109},
archivePrefix = {arXiv},
       eprint = {1107.5825},
 primaryClass = {astro-ph.GA},
       adsurl = {https://ui.adsabs.harvard.edu/abs/2011ApJ...740..109O}
}

@ARTICLE{2020SciPy-NMeth,
  author  = {Virtanen, Pauli and Gommers, Ralf and Oliphant, Travis E. and
            Haberland, Matt and Reddy, Tyler and Cournapeau, David and
            Burovski, Evgeni and Peterson, Pearu and Weckesser, Warren and
            Bright, Jonathan and {van der Walt}, St{\'e}fan J. and
            Brett, Matthew and Wilson, Joshua and Millman, K. Jarrod and
            Mayorov, Nikolay and Nelson, Andrew R. J. and Jones, Eric and
            Kern, Robert and Larson, Eric and Carey, C J and
            Polat, {\.I}lhan and Feng, Yu and Moore, Eric W. and
            {VanderPlas}, Jake and Laxalde, Denis and Perktold, Josef and
            Cimrman, Robert and Henriksen, Ian and Quintero, E. A. and
            Harris, Charles R. and Archibald, Anne M. and
            Ribeiro, Ant{\^o}nio H. and Pedregosa, Fabian and
            {van Mulbregt}, Paul and {SciPy 1.0 Contributors}},
  title   = {{{SciPy} 1.0: Fundamental Algorithms for Scientific
            Computing in Python}},
  journal = {    NaMet     },
  year    = {2020},
  volume  = {17},
  pages   = {261--272},
  adsurl  = {https://rdcu.be/b08Wh},
  doi     = {10.1038/s41592-019-0686-2},
}

@ARTICLE{2007WRR....43.3423V,
       author = {{Vugrin}, K.~W. and {Swiler}, L.~P. and {Roberts}, R.~M. and {Stucky-Mack}, N.~J. and {Sullivan}, S.~P.},
        title = "{Confidence region estimation techniques for nonlinear regression in groundwater flow: Three case studies}",
      journal = {    WRR        },
         year = 2007,
        month = mar,
       volume = {43},
       number = {3},
          eid = {W03423},
        pages = {W03423},
          doi = {10.1029/2005WR004804},
       adsurl = {https://ui.adsabs.harvard.edu/abs/2007WRR....43.3423V}
}

@ARTICLE{Molpeceres_ea2025,
       author = {{Molpeceres}, Germ{\'a}n and {Enrique-Romero}, Joan and {Ishibashi}, Atsuki and {Oba}, Yasuhiro and {Hidaka}, Hiroshi and {Lamberts}, Thanja and {Aikawa}, Yuri and {Watanabe}, Naoki},
        title = "{Hydrogenation of HOCO and formation of interstellar CO$_{2}$: a not so straightforward relation}",
      journal = {\mnras},
         year = 2025,
        month = apr,
       volume = {538},
       number = {3},
        pages = {1565-1575},
          doi = {10.1093/mnras/staf383},
archivePrefix = {arXiv},
       eprint = {2503.01692},
 primaryClass = {astro-ph.GA},
       adsurl = {https://ui.adsabs.harvard.edu/abs/2025MNRAS.538.1565M}
}

@ARTICLE{Borshcheva_ea2025,
       author = {{Borshcheva}, Katerina and {Fedoseev}, Gleb and {Punanova}, Anna F. and {Caselli}, Paola and {Jim{\'e}nez-Serra}, Izaskun and {Vasyunin}, Anton I.},
        title = "{Formation of Complex Organic Molecules in Prestellar Cores: The Role of Nondiffusive Grain Chemistry}",
      journal = {\apj},
         year = 2025,
        month = sep,
       volume = {990},
       number = {2},
          eid = {163},
        pages = {163},
          doi = {10.3847/1538-4357/adea73},
archivePrefix = {arXiv},
       eprint = {2507.00338},
 primaryClass = {astro-ph.GA},
       adsurl = {https://ui.adsabs.harvard.edu/abs/2025ApJ...990..163B}
}

@ARTICLE{Song_Kaestner2016,
       author = {{Song}, Lei and {K{\"a}stner}, Johannes},
        title = "{Formation of the prebiotic molecule NH2CHO on astronomical amorphous solid water surfaces: accurate tunneling rate calculations}",
      journal = {PCCP)},
         year = 2016,
        month = jan,
       volume = {18},
       number = {42},
        pages = {29278-29285},
          doi = {10.1039/C6CP05727F},
archivePrefix = {arXiv},
       eprint = {1610.01007},
 primaryClass = {physics.chem-ph},
       adsurl = {https://ui.adsabs.harvard.edu/abs/2016PCCP...1829278S}
}

@ARTICLE{Rawlings_ea1992,
       author = {{Rawlings}, J.~M.~C. and {Hartquist}, T.~W. and {Menten}, K.~M. and {Williams}, D.~A.},
        title = "{Direct diagnosis of infall in collapsing protostars - I. The theoretical identification of molecular species with broad velocity distributions.}",
      journal = {\mnras},
         year = 1992,
        month = apr,
       volume = {255},
        pages = {471-485},
          doi = {10.1093/mnras/255.3.471},
       adsurl = {https://ui.adsabs.harvard.edu/abs/1992MNRAS.255..471R}
}

@ARTICLE{Jin_Garrod2020,
       author = {{Jin}, Mihwa and {Garrod}, Robin T.},
        title = "{Formation of Complex Organic Molecules in Cold Interstellar Environments through Nondiffusive Grain-surface and Ice-mantle Chemistry}",
      journal = {\apjs},
         year = 2020,
        month = aug,
       volume = {249},
       number = {2},
          eid = {26},
        pages = {26},
          doi = {10.3847/1538-4365/ab9ec8},
archivePrefix = {arXiv},
       eprint = {2006.11127},
 primaryClass = {astro-ph.GA},
       adsurl = {https://ui.adsabs.harvard.edu/abs/2020ApJS..249...26J}
}

@ARTICLE{Jorgensen_ea2009,
       author = {{J{\o}rgensen}, J.~K. and {van Dishoeck}, E.~F. and {Visser}, R. and {Bourke}, T.~L. and {Wilner}, D.~J. and {Lommen}, D. and {Hogerheijde}, M.~R. and {Myers}, P.~C.},
        title = "{PROSAC: a submillimeter array survey of low-mass protostars. II. The mass evolution of envelopes, disks, and stars from the Class 0 through I stages}",
      journal = {\aap},
         year = 2009,
        month = nov,
       volume = {507},
       number = {2},
        pages = {861-879},
          doi = {10.1051/0004-6361/200912325},
archivePrefix = {arXiv},
       eprint = {0909.3386},
 primaryClass = {astro-ph.SR},
       adsurl = {https://ui.adsabs.harvard.edu/abs/2009A&A...507..861J}
}

@ARTICLE{Hocuk_ea2017,
       author = {{Hocuk}, S. and {Sz{\H{u}}cs}, L. and {Caselli}, P. and {Cazaux}, S. and {Spaans}, M. and {Esplugues}, G.~B.},
        title = "{Parameterizing the interstellar dust temperature}",
      journal = {\aap},
         year = 2017,
        month = aug,
       volume = {604},
          eid = {A58},
        pages = {A58},
          doi = {10.1051/0004-6361/201629944},
archivePrefix = {arXiv},
       eprint = {1704.02763},
 primaryClass = {astro-ph.GA},
       adsurl = {https://ui.adsabs.harvard.edu/abs/2017A&A...604A..58H}
}

@ARTICLE{Garrod_Pauly2011,
       author = {{Garrod}, R.~T. and {Pauly}, T.},
        title = "{On the Formation of CO$_{2}$ and Other Interstellar Ices}",
      journal = {\apj},
         year = 2011,
        month = jul,
       volume = {735},
       number = {1},
          eid = {15},
        pages = {15},
          doi = {10.1088/0004-637X/735/1/15},
archivePrefix = {arXiv},
       eprint = {1106.0540},
 primaryClass = {astro-ph.GA},
       adsurl = {https://ui.adsabs.harvard.edu/abs/2011ApJ...735...15G}
}

@ARTICLE{Garrod_ea2022,
       author = {{Garrod}, Robin T. and {Jin}, Miwha and {Matis}, Kayla A. and {Jones}, Dylan and {Willis}, Eric R. and {Herbst}, Eric},
        title = "{Formation of Complex Organic Molecules in Hot Molecular Cores through Nondiffusive Grain-surface and Ice-mantle Chemistry}",
      journal = {\apjs},
         year = 2022,
        month = mar,
       volume = {259},
       number = {1},
          eid = {1},
        pages = {1},
          doi = {10.3847/1538-4365/ac3131},
archivePrefix = {arXiv},
       eprint = {2110.09743},
 primaryClass = {astro-ph.GA},
       adsurl = {https://ui.adsabs.harvard.edu/abs/2022ApJS..259....1G}
}

@ARTICLE{Petrashkevich_ea2026,
       author = {{Petrashkevich}, Igor and {Pavlyuchenkov}, Yaroslav and {Punanova}, Anna and {Ozhiganov}, Maksim and {Nakibov}, Ruslan and {Karteyeva}, Varvara and {Salii}, Svetlana and {Sobolev}, Andrej and {Medvedev}, Mikhail and {Vasyunin}, Anton},
        title = "{Mapping Interstellar Ice Inventory toward Class 0 Protostars in Star-forming Region Orion A with JWST Data}",
      journal = {\apj},
         year = 2026,
        month = jul,
       volume = {1006},
       number = {1},
          eid = {46},
        pages = {46},
          doi = {10.3847/1538-4357/ae7b20},
archivePrefix = {arXiv},
       eprint = {2606.09776},
 primaryClass = {astro-ph.GA},
       adsurl = {https://ui.adsabs.harvard.edu/abs/2026ApJ..1006...46P}
}

@ARTICLE{Jamieson_ea2005,
       author = {{Jamieson}, Corey S. and {Bennett}, Chris J. and {Mebel}, Alexander M. and {Kaiser}, Ralf I.},
        title = "{Investigating the Mechanism for the Formation of Nitrous Oxide [N$_{2}$O(X $^{1}${\ensuremath{\Sigma}}$^{+}$)] in Extraterrestrial Ices}",
      journal = {\apj},
         year = 2005,
        month = may,
       volume = {624},
       number = {1},
        pages = {436-447},
          doi = {10.1086/428933},
       adsurl = {https://ui.adsabs.harvard.edu/abs/2005ApJ...624..436J}
}

@ARTICLE{2008ApJ...678.1005P,
       author = {{Pontoppidan}, Klaus M. and {Boogert}, A.~C.~A. and {Fraser}, Helen J. and {van Dishoeck}, Ewine F. and {Blake}, Geoffrey A. and {Lahuis}, Fred and {{\"O}berg}, Karin I. and {Evans}, II, Neal J. and {Salyk}, Colette},
        title = "{The c2d Spitzer Spectroscopic Survey of Ices around Low-Mass Young Stellar Objects. II. CO$_{2}$}",
      journal = {\apj},
         year = 2008,
        month = may,
       volume = {678},
       number = {2},
        pages = {1005-1031},
          doi = {10.1086/533431},
archivePrefix = {arXiv},
       eprint = {0711.4616},
 primaryClass = {astro-ph},
       adsurl = {https://ui.adsabs.harvard.edu/abs/2008ApJ...678.1005P}
}

@ARTICLE{Clement_ea23,
       author = {{Cl{\'e}ment}, A. and {Taillard}, A. and {Wakelam}, V. and {Gratier}, P. and {Loison}, J.-C. and {Dartois}, E. and {Dulieu}, F. and {Noble}, J.~A. and {Chabot}, M.},
        title = "{Astrochemical models of interstellar ices: History matters}",
      journal = {\aap},
         year = 2023,
        month = jul,
       volume = {675},
          eid = {A165},
        pages = {A165},
          doi = {10.1051/0004-6361/202346188},
archivePrefix = {arXiv},
       eprint = {2306.08346},
 primaryClass = {astro-ph.GA},
       adsurl = {https://ui.adsabs.harvard.edu/abs/2023A&A...675A.165C}
}

@ARTICLE{JimenezSerra_ea25,
       author = {{Jim{\'e}nez-Serra}, Izaskun and {Meg{\'\i}as}, Andr{\'e}s and {Salaris}, Joseph and {Cuppen}, Herma and {Taillard}, Ang{\`e}le and {Jin}, Miwha and {Wakelam}, Valentine and {Vasyunin}, Anton I. and {Caselli}, Paola and {Pendleton}, Yvonne J. and {Dartois}, Emmanuel and {Noble}, Jennifer A. and {Viti}, Serena and {Borshcheva}, Katerina and {Garrod}, Robin T. and {Lamberts}, Thanja and {Fraser}, Helen and {Melnick}, Gary and {McClure}, Melissa and {Rocha}, Will and {Drozdovskaya}, Maria N. and {Lis}, Dariusz C.},
        title = "{Modelling methanol and hydride formation in the JWST Ice Age era}",
      journal = {\aap},
         year = 2025,
        month = mar,
       volume = {695},
          eid = {A247},
        pages = {A247},
          doi = {10.1051/0004-6361/202452389},
archivePrefix = {arXiv},
       eprint = {2502.10123},
 primaryClass = {astro-ph.GA},
       adsurl = {https://ui.adsabs.harvard.edu/abs/2025A&A...695A.247J}
}

@ARTICLE{2022AA...665A..89R,
       author = {{Rachid}, M.~G. and {Rocha}, W.~R.~M. and {Linnartz}, H.},
        title = "{Infrared spectra of complex organic molecules in astronomically relevant ice mixtures. V. Methyl cyanide (acetonitrile)}",
      journal = {\aap},
         year = 2022,
        month = sep,
       volume = {665},
          eid = {A89},
        pages = {A89},
          doi = {10.1051/0004-6361/202243417},
archivePrefix = {arXiv},
       eprint = {2207.12502},
 primaryClass = {astro-ph.GA},
       adsurl = {https://ui.adsabs.harvard.edu/abs/2022A&A...665A..89R}
}

@ARTICLE{2024A&A...686A..71N,
       author = {{Nazari}, P. and {Rocha}, W.~R.~M. and {Rubinstein}, A.~E. and {Slavicinska}, K. and {Rachid}, M.~G. and {van Dishoeck}, E.~F. and {Megeath}, S.~T. and {Gutermuth}, R. and {Tyagi}, H. and {Brunken}, N. and {Narang}, M. and {Manoj}, P. and {Watson}, D.~M. and {Evans}, N.~J. and {Federman}, S. and {Muzerolle Page}, J. and {Anglada}, G. and {Beuther}, H. and {Klaassen}, P. and {Looney}, L.~W. and {Osorio}, M. and {Stanke}, T. and {Yang}, Y.-L.},
        title = "{Hunting for complex cyanides in protostellar ices with the JWST. A tentative detection of CH$_{3}$CN and C$_{2}$H$_{5}$CN}",
      journal = {\aap},
         year = 2024,
        month = jun,
       volume = {686},
          eid = {A71},
        pages = {A71},
          doi = {10.1051/0004-6361/202348695},
archivePrefix = {arXiv},
       eprint = {2401.07901},
 primaryClass = {astro-ph.GA},
       adsurl = {https://ui.adsabs.harvard.edu/abs/2024A&A...686A..71N}
}

@ARTICLE{2017A&A...601A..49B,
       author = {{Belloche}, A. and {Meshcheryakov}, A.~A. and {Garrod}, R.~T. and {Ilyushin}, V.~V. and {Alekseev}, E.~A. and {Motiyenko}, R.~A. and {Margul{\`e}s}, L. and {M{\"u}ller}, H.~S.~P. and {Menten}, K.~M.},
        title = "{Rotational spectroscopy, tentative interstellar detection, and chemical modeling of N-methylformamide}",
      journal = {\aap},
         year = 2017,
        month = may,
       volume = {601},
          eid = {A49},
        pages = {A49},
          doi = {10.1051/0004-6361/201629724},
archivePrefix = {arXiv},
       eprint = {1701.04640},
 primaryClass = {astro-ph.GA},
       adsurl = {https://ui.adsabs.harvard.edu/abs/2017A&A...601A..49B}
}

@ARTICLE{2009A&A...499..215B,
       author = {{Belloche}, A. and {Garrod}, R.~T. and {M{\"u}ller}, H.~S.~P. and {Menten}, K.~M. and {Comito}, C. and {Schilke}, P.},
        title = "{Increased complexity in interstellar chemistry: detection and chemical modeling of ethyl formate and n-propyl cyanide in Sagittarius B2(N)}",
      journal = {\aap},
         year = 2009,
        month = may,
       volume = {499},
       number = {1},
        pages = {215-232},
          doi = {10.1051/0004-6361/200811550},
archivePrefix = {arXiv},
       eprint = {0902.4694},
 primaryClass = {astro-ph.GA},
       adsurl = {https://ui.adsabs.harvard.edu/abs/2009A&A...499..215B}
}

@ARTICLE{2008ApJ...682..283G,
       author = {{Garrod}, Robin T. and {Widicus Weaver}, Susanna L. and {Herbst}, Eric},
        title = "{Complex Chemistry in Star-forming Regions: An Expanded Gas-Grain Warm-up Chemical Model}",
      journal = {\apj},
         year = 2008,
        month = jul,
       volume = {682},
       number = {1},
        pages = {283-302},
          doi = {10.1086/588035},
archivePrefix = {arXiv},
       eprint = {0803.1214},
 primaryClass = {astro-ph},
       adsurl = {https://ui.adsabs.harvard.edu/abs/2008ApJ...682..283G}
}

@ARTICLE{2026A&A...711A.240N,
       author = {{Nazari}, P. and {Brunken}, N. and {Chen}, Y. and {Slavicinska}, K. and {van Dishoeck}, E.~F. and {Rocha}, W.~R.~M. and {Boogert}, A.~C.~A. and {Navarro}, M.~G. and {Le Gouellec}, V.~J.~M. and {Francis}, L. and {Tychoniec}, L. and {Caratti o Garatti}, A. and {Gieser}, C. and {Greene}, T.~P. and {Kavanagh}, P.~J.},
        title = "{JOYS+: Analyses of OCN$^{-}$, N$_{2}$O, NO, and complex cyanides in ices: Thermal processing results in a modest increase of OCN$^{{\ensuremath{-}}}$ ice}",
      journal = {\aap},
         year = 2026,
        month = jul,
       volume = {711},
          eid = {A240},
        pages = {A240},
          doi = {10.1051/0004-6361/202557830},
archivePrefix = {arXiv},
       eprint = {2604.25915},
 primaryClass = {astro-ph.GA},
       adsurl = {https://ui.adsabs.harvard.edu/abs/2026A&A...711A.240N}
}

@BOOK{jabi,
       author = {{Burnham}, K.~P. and {Anderson}, D.~R.},
        title = "{Model Selection and Multimodel Inference: A Practical Information-Theoretic Approach}",
    publisher = {Springer},
      address = {New York},
         year = 2002,
      edition = {2nd},
          doi = {10.1007/b97636},
         isbn = {978-0-387-95364-9}
}

@ARTICLE{2024NatAs...8..359D,
       author = {{Dartois}, E. and {Noble}, J.~A. and {Caselli}, P. and {Fraser}, H.~J. and {Jim{\'e}nez-Serra}, I. and {Mat{\'e}}, B. and {McClure}, M.~K. and {Melnick}, G.~J. and {Pendleton}, Y.~J. and {Shimonishi}, T. and {Smith}, Z.~L. and {Sturm}, J.~A. and {Taillard}, A. and {Wakelam}, V. and {Boogert}, A.~C.~A. and {Drozdovskaya}, M.~N. and {Erkal}, J. and {Harsono}, D. and {Herrero}, V.~J. and {Ioppolo}, S. and {Linnartz}, H. and {McGuire}, B.~A. and {Perotti}, G. and {Qasim}, D. and {Rocha}, W.~R.~M.},
        title = "{Spectroscopic sizing of interstellar icy grains with JWST}",
      journal = {NatAs},
         year = 2024,
        month = mar,
       volume = {8},
        pages = {359-367},
          doi = {10.1038/s41550-023-02155-x},
       adsurl = {https://ui.adsabs.harvard.edu/abs/2024NatAs...8..359D}
}

@article{jabi2,
    author = {Kenneth P. Burnham and David R. Anderson},
    title ={Multimodel Inference: Understanding AIC and BIC in Model Selection},
    journal = {Sociological Methods \& Research},
    volume = {33},
    number = {2},
    pages = {261-304},
    year = {2004},
    doi = {10.1177/0049124104268644},
    URL = {https://doi.org/10.1177/0049124104268644},
    eprint = {https://doi.org/10.1177/0049124104268644}
}

@ARTICLE{Gerakines_ea2004,
       author = {{Gerakines}, P.~A. and {Moore}, M.~H. and {Hudson}, R.~L.},
        title = "{Ultraviolet photolysis and proton irradiation of astrophysical ice analogs containing hydrogen cyanide}",
      journal = {\icarus},
         year = 2004,
        month = jul,
       volume = {170},
       number = {1},
        pages = {202-213},
          doi = {10.1016/j.icarus.2004.02.005},
       adsurl = {https://ui.adsabs.harvard.edu/abs/2004Icar..170..202G}
}

@ARTICLE{1974ITAC...19..716A,
       author = {{Akaike}, H.},
        title = "{A New Look at the Statistical Model Identification}",
      journal = {ITAC},
         year = 1974,
        month = jan,
       volume = {19},
        pages = {716-723},
          doi = {10.1109/TAC.1974.1100705},
       adsurl = {https://ui.adsabs.harvard.edu/abs/1974ITAC...19..716A}
}

@article{Sugiura01011978,
author = {Nariaki Sugiura},
title = {Further analysis of the data by Akaike's information criterion and the finite corrections},
journal = {Communications in Statistics - Theory and Methods},
volume = {7},
number = {1},
pages = {13--26},
year = {1978},
publisher = {Taylor \& Francis},
doi = {10.1080/03610927808827599},
URL = {https://doi.org/10.1080/03610927808827599},
eprint = {https://doi.org/10.1080/03610927808827599}
}

@article{10.1093/biomet/76.2.297,
    author = {Hurvich, CLIFFORD M. and Tsai, CHIH-LING},
    title = {Regression and time series model selection in small samples},
    journal = {Biometrika},
    volume = {76},
    number = {2},
    pages = {297-307},
    year = {1989},
    month = {06},
    issn = {0006-3444},
    doi = {10.1093/biomet/76.2.297},
    url = {https://doi.org/10.1093/biomet/76.2.297},
    eprint = {https://academic.oup.com/biomet/article-pdf/76/2/297/737009/76-2-297.pdf},
}

@ARTICLE{1997JChPh.107.9764B,
       author = {{Brown}, Steven S. and {Berghout}, H. Laine and {Crim}, F. Fleming},
        title = "{Raman spectroscopy of the N-C-O symmetric ({\ensuremath{\nu}}$_{3}$) and antisymmetric ({\ensuremath{\nu}}$_{2}$) stretch fundamentals in HNCO}",
      journal = {\jcp},
         year = 1997,
        month = dec,
       volume = {107},
       number = {23},
        pages = {9764-9771},
          doi = {10.1063/1.475274},
       adsurl = {https://ui.adsabs.harvard.edu/abs/1997JChPh.107.9764B}
}

@ARTICLE{2024A&A...690A.205C,
       author = {{Chen}, Y. and {Rocha}, W.~R.~M. and {van Dishoeck}, E.~F. and {van Gelder}, M.~L. and {Nazari}, P. and {Slavicinska}, K. and {Francis}, L. and {Tabone}, B. and {Ressler}, M.~E. and {Klaassen}, P.~D. and {Beuther}, H. and {Boogert}, A.~C.~A. and {Gieser}, C. and {Kavanagh}, P.~J. and {Perotti}, G. and {Le Gouellec}, V.~J.~M. and {Majumdar}, L. and {G{\"u}del}, M. and {Henning}, Th.},
        title = "{JOYS+: The link between the ice and gas of complex organic molecules: Comparing JWST and ALMA data of two low-mass protostars}",
      journal = {\aap},
         year = 2024,
        month = oct,
       volume = {690},
          eid = {A205},
        pages = {A205},
          doi = {10.1051/0004-6361/202450706},
archivePrefix = {arXiv},
       eprint = {2407.20066},
 primaryClass = {astro-ph.GA},
       adsurl = {https://ui.adsabs.harvard.edu/abs/2024A&A...690A.205C}
}

@article{Stephens1974EDFSF,
  title={EDF Statistics for Goodness of Fit and Some Comparisons},
  author={Michael A. Stephens},
  journal={Journal of the American Statistical Association},
  year={1974},
  volume={69},
  pages={730-737},
  url={https://api.semanticscholar.org/CorpusID:122607594},
  doi={https://doi.org/10.2307/2286009}
}

@BOOK{2012msma.book.....F,
       author = {{Feigelson}, Eric D. and {Babu}, G. Jogesh},
        title = "{Modern Statistical Methods for Astronomy}",
         year = 2012,
          doi = {10.48550/arXiv.1205.2064},
       adsurl = {https://ui.adsabs.harvard.edu/abs/2012msma.book.....F}
}

@ARTICLE{Tobin2016,
       author = {{Tobin}, John J. and {Looney}, Leslie W. and {Li}, Zhi-Yun and {Chandler}, Claire J. and {Dunham}, Michael M. and {Segura-Cox}, Dominique and {Sadavoy}, Sarah I. and {Melis}, Carl and {Harris}, Robert J. and {Kratter}, Kaitlin and {Perez}, Laura},
        title = "{The VLA Nascent Disk and Multiplicity Survey of Perseus Protostars (VANDAM). II. Multiplicity of Protostars in the Perseus Molecular Cloud}",
      journal = {\apj},
         year = 2016,
        month = feb,
       volume = {818},
       number = {1},
          eid = {73},
        pages = {73},
          doi = {10.3847/0004-637X/818/1/73},
archivePrefix = {arXiv},
       eprint = {1601.00692},
 primaryClass = {astro-ph.SR},
       adsurl = {https://ui.adsabs.harvard.edu/abs/2016ApJ...818...73T}
}

@ARTICLE{2016ApJS..224....5F,
       author = {{Furlan}, E. and {Fischer}, W.~J. and {Ali}, B. and {Stutz}, A.~M. and {Stanke}, T. and {Tobin}, J.~J. and {Megeath}, S.~T. and {Osorio}, M. and {Hartmann}, L. and {Calvet}, N. and {Poteet}, C.~A. and {Booker}, J. and {Manoj}, P. and {Watson}, D.~M. and {Allen}, L.},
        title = "{The Herschel Orion Protostar Survey: Spectral Energy Distributions and Fits Using a Grid of Protostellar Models}",
      journal = {\apjs},
         year = 2016,
        month = may,
       volume = {224},
       number = {1},
          eid = {5},
        pages = {5},
          doi = {10.3847/0067-0049/224/1/5},
archivePrefix = {arXiv},
       eprint = {1602.07314},
 primaryClass = {astro-ph.SR},
       adsurl = {https://ui.adsabs.harvard.edu/abs/2016ApJS..224....5F}
}

@ARTICLE{Ortiz-Le2018,
       author = {{Ortiz-Le{\'o}n}, Gisela N. and {Loinard}, Laurent and {Dzib}, Sergio A. and {Galli}, Phillip A.~B. and {Kounkel}, Marina and {Mioduszewski}, Amy J. and {Rodr{\'\i}guez}, Luis F. and {Torres}, Rosa M. and {Hartmann}, Lee and {Boden}, Andrew F. and {Evans}, II, Neal J. and {Brice{\~n}o}, Cesar and {Tobin}, John J.},
        title = "{The Gould{\textquoteright}s Belt Distances Survey (GOBELINS). V. Distances and Kinematics of the Perseus Molecular Cloud}",
      journal = {\apj},
         year = 2018,
        month = sep,
       volume = {865},
       number = {1},
          eid = {73},
        pages = {73},
          doi = {10.3847/1538-4357/aada49},
archivePrefix = {arXiv},
       eprint = {1808.03499},
 primaryClass = {astro-ph.SR},
       adsurl = {https://ui.adsabs.harvard.edu/abs/2018ApJ...865...73O}
}

@ARTICLE{2016ApJ...823..125C,
       author = {{Chen}, Huei-Ru Vivien and {Keto}, Eric and {Zhang}, Qizhou and {Sridharan}, T.~K. and {Liu}, Sheng-Yuan and {Su}, Yu-Nung},
        title = "{A Hot and Massive Accretion Disk around the High-mass Protostar IRAS 20126+4104}",
     journal = {ApJ},
         year = 2016,
        month = jun,
       volume = {823},
       number = {2},
          eid = {125},
        pages = {125},
          doi = {10.3847/0004-637X/823/2/125},
archivePrefix = {arXiv},
       eprint = {1604.00523},
 primaryClass = {astro-ph.GA},
       adsurl = {https://ui.adsabs.harvard.edu/abs/2016ApJ...823..125C}
}

@ARTICLE{2024A&A...683A.124R,
       author = {{Rocha}, W.~R.~M. and {van Dishoeck}, E.~F. and {Ressler}, M.~E. and {van Gelder}, M.~L. and {Slavicinska}, K. and {Brunken}, N.~G.~C. and {Linnartz}, H. and {Ray}, T.~P. and {Beuther}, H. and {Caratti o Garatti}, A. and {Geers}, V. and {Kavanagh}, P.~J. and {Klaassen}, P.~D. and {Justtanont}, K. and {Chen}, Y. and {Francis}, L. and {Gieser}, C. and {Perotti}, G. and {Tychoniec}, {\L}. and {Barsony}, M. and {Majumdar}, L. and {le Gouellec}, V.~J.~M. and {Chu}, L.~E.~U. and {Lew}, B.~W.~P. and {Henning}, Th. and {Wright}, G.},
        title = "{JWST Observations of Young protoStars (JOYS+): Detecting icy complex organic molecules and ions. I. CH$_{4}$, SO$_{2}$, HCOO$^{‒}$, OCN$^{‒}$, H$_{2}$CO, HCOOH, CH$_{3}$CH$_{2}$OH, CH$_{3}$CHO, CH$_{3}$OCHO, and CH$_{3}$COOH}",
      journal = {\aap},
         year = 2024,
        month = mar,
       volume = {683},
          eid = {A124},
        pages = {A124},
          doi = {10.1051/0004-6361/202348427},
archivePrefix = {arXiv},
       eprint = {2312.06834},
 primaryClass = {astro-ph.SR},
       adsurl = {https://ui.adsabs.harvard.edu/abs/2024A&A...683A.124R}
}

@ARTICLE{MD24,
       author = {{Mart{\'\i}n-Dom{\'e}nech}, Rafael and {DelFranco}, Alexander and {{\"O}berg}, Karin I. and {Rajappan}, Mahesh},
        title = "{Organic Chemistry in the H$_{2}$-bearing, CO-rich Interstellar Ice Layer at Temperatures Relevant to Dense Cloud Interiors}",
      journal = {\apj},
         year = 2024,
        month = feb,
       volume = {962},
       number = {2},
          eid = {107},
        pages = {107},
          doi = {10.3847/1538-4357/ad187e},
archivePrefix = {arXiv},
       eprint = {2401.04024},
 primaryClass = {astro-ph.GA},
       adsurl = {https://ui.adsabs.harvard.edu/abs/2024ApJ...962..107M}
}

@ARTICLE{MD20,
       author = {{Mart{\'\i}n-Dom{\'e}nech}, Rafael and {Maksiutenko}, Pavlo and {{\"O}berg}, Karin I. and {Rajappan}, Mahesh},
        title = "{Exploring the Chemistry Induced by Energetic Processing of the H2-bearing, CO-rich Apolar Ice Layer}",
      journal = {\apj},
         year = 2020,
        month = oct,
       volume = {902},
       number = {2},
          eid = {116},
        pages = {116},
          doi = {10.3847/1538-4357/abb59f},
archivePrefix = {arXiv},
       eprint = {2009.01335},
 primaryClass = {astro-ph.GA},
       adsurl = {https://ui.adsabs.harvard.edu/abs/2020ApJ...902..116M}
}
\bibliographystyle{aasjournal}

\end{document}